\documentclass[fleqn,usenatbib]{mnras}

\usepackage{newtxtext,newtxmath}

\usepackage[T1]{fontenc}

\DeclareRobustCommand{\VAN}[3]{#2}
\let\VANthebibliography\thebibliography
\def\thebibliography{\DeclareRobustCommand{\VAN}[3]{##3}\VANthebibliography}

\newcommand{\rt}[1]{\textcolor{black}{#1}}

\usepackage{graphicx}	
\usepackage{amsmath}	
\usepackage{subfig}

\title[Exoplanet Discovery with the GPFC Method]{The GPU Phase Folding and Deep Learning Method for Detecting Exoplanet Transits}

\author[K. Wang et al.]{
Kaitlyn Wang$^{1,2}\thanks{E-mail:24kaitlynw@students.harker.org},$
Jian Ge$^{3}\thanks{E-mail:jge@shao.ac.cn},$
Kevin Willis$^{2},$
Kevin Wang$^{4},$
Yinan Zhao$^{5}$
\\
$^{1}$The Harker School, 500 Saratoga Ave, San Jose, CA 95129, USA\\
$^{2}$Science Talent Training Center, Gainesville, FL, 32606 USA\\
$^{3}$Shanghai Astronomical Observatory, Shanghai 200030, China\\
$^{4}$Princeton University, PO Box 430 Princeton, NJ 08544, USA\\
$^{5}$Department of Astronomy, University of Geneva, Switzerland
}

\date{Accepted 2024 January 20. Received 2024 January 20; in original form 2023 December 04}

\pubyear{2024}

\begin{document}
\label{firstpage}
\pagerange{\pageref{firstpage}--\pageref{lastpage}}
\maketitle

\begin{abstract}
This paper presents GPFC, a novel Graphics Processing Unit (GPU) Phase Folding and Convolutional Neural Network (CNN) system to detect exoplanets using the transit method. We devise a fast folding algorithm parallelized on a GPU to amplify low signal-to-noise ratio transit signals, allowing a search at high precision and speed. A CNN trained on two million synthetic light curves reports a score indicating the likelihood of a planetary signal at each period. \rt{While the GPFC method has broad applicability across period ranges, this research specifically focuses on detecting ultra-short-period planets with orbital periods less than one day.} GPFC improves on speed by three orders of magnitude over the predominant Box-fitting Least Squares (BLS) method. Our simulation results show GPFC achieves 97\% training accuracy, higher true positive rate at the same false positive rate of detection, and higher precision at the same recall rate when compared to BLS. GPFC recovers 100\% of known ultra-short-period planets in \textit{Kepler} light curves from a blind search. These results highlight the promise of GPFC as an alternative approach to the traditional BLS algorithm for finding new transiting exoplanets in data taken with \textit{Kepler} and other space transit missions such as K2, TESS and future PLATO and Earth 2.0.

\end{abstract}

\begin{keywords}
techniques: photometric -- methods: data analysis -- planets and satellites: detection --  catalogues -- surveys
\end{keywords}



\section{Introduction}

Since the discovery of the first exoplanets \citep{wolszczan_frail_1992, major_queloz_1995}, more than 5,000 exoplanets have been found and many thousands of candidates have yet to be confirmed. Compared with the other major exoplanet detection methods---radial velocity \citep{campbell_rv_1988}, direct imaging \citep{Chauvin_2004}, and gravitational microlensing \citep{Beaulieu_2006}---the transit method \citep{Charbonneau_2000} has made the biggest contribution empowered by the large scale transit surveys including \textit{Kepler} \citep{borucki_2010}, K2 \citep{Howell_2014}, TESS \citep{Ricker_2014} and beyond.

At a high level, planetary transit detection involves the following general steps. First, the light curve data is pre-processed and detrended to remove stellar variability \citep{Stumpe_2012, Smith_2012}. After that, a variety of algorithms are used to search for periodic transit signals in the light curve. The Box-fitting Least Square (BLS) method, introduced by \citet{Kov_cs_2002}, has been the predominant method for transit searches in large data sets and is widely used in the ground-based and space-based surveys. There have been various extensions and optimizations on the original BLS method: \citep{Ofir_2014, Carter_2013, Boufleur_2013, Renner_2008, Collier_Cameron_2006, Hartman_2016, Caceres_2019}. \citet{Panahi_2021} optimized BLS for low-cadence surveys such as Gaia \citep{Gaia_Collab_2016}, and \citet{Shahaf_2022} proposed an efficient periodicity detection algorithm by combining two long withstanding techniques - the fast-folding algorithm (FFA: \citep{Staelin_1969}) and BLS. Following transit signal detection, a list of threshold crossing events (TCEs) is generated. Then, a vetting process is conducted to filter out a vast number of false positives in the TCEs caused by instrumental noise or astrophysical variability. Various machine learning auto-vetting methods have been developed, including Robovetter \citep{Thompson_2018}, Autovetter \citep{McCauliff_2015} and Astronet \citep{Shallue_2018}. Astronet employed deep learning (convolutional neural network) to vet \textit{Kepler} candidates, and it was thereafter adapted to more surveys such as K2 \citep{Dattilo_2019}, NGTS \citep{Chaushev_2019},  WASP \citep{Schanche_2018} and TESS \citep{Yu_2019, Osborn_2020, Ofman_2022, Olmschenk_2021, Rao_2021}.

Meanwhile, some researchers have been exploring a different approach which detects exoplanets directly from light curves via machine learning without the involvement of the BLS method \citep{Pearson_2017, Zucker_2018, Chintarungruangchai_2019, Malik_2021, Cui_2021}. Among these, some utilize the phase folding technique whereas others do not. In general, the methods processing light curves without phase folding limit their sensitivity to transit signals with signal-to-noise ratios (SNR) above 10; for the methods that involve phase folding \citep{Pearson_2017, Yeh_2020}, since the resolution of the trial folding periods limits the accuracy of unknown transit detection, those efforts mainly focused on simulated data rather than applying their neural networks to search for new candidates. Practically, in order to detect shallow transit signals generated by small exoplanets, there must be a high resolution of trial folding periods, which converts to prohibitive computation time. This is the critical problem that motivates our novel method, the GPU (Graphics Processing Units) Phase Folding and Convolutional Neural Network (GPFC) method. With GPFC, we increase computational speed to achieve phase folding at high resolution trial periods. In GPFC we developed a scalable phase folding algorithm leveraging GPU's parallelism to process phase folding with high precision, together with a convolutional neural network (CNN) to evaluate transit signals from the high-dimensional folded results. We demonstrate that the GPFC detection system is capable of searching exoplanets in large volume of \textit{Kepler} data at three orders of magnitude higher speed than traditional BLS, and it also reports new exoplanet candidates undetected in previous research work.

The organization of this paper is as follows: Section 2 delineates the foundational principles of the GPFC method, detailing each component within its operational pipeline. Section 3 delves into a detailed description of the simulation tests we employed to compare the GPFC method and the classic BLS method. In Section 4, we demonstrate that the GPFC detection system can recover all of confirmed USPs in the \textit{Kepler} Archive--a validation for its potential for new exoplanetary discoveries. Finally, in Section 5, we present a discussion of our method, comparison with various implementations of the BLS method, and outline prospective directions for future research.

\section{Methods}

\subsection{Overview of the GPFC Method}

\begin{figure}
    \centering
   \includegraphics[scale = 0.25]{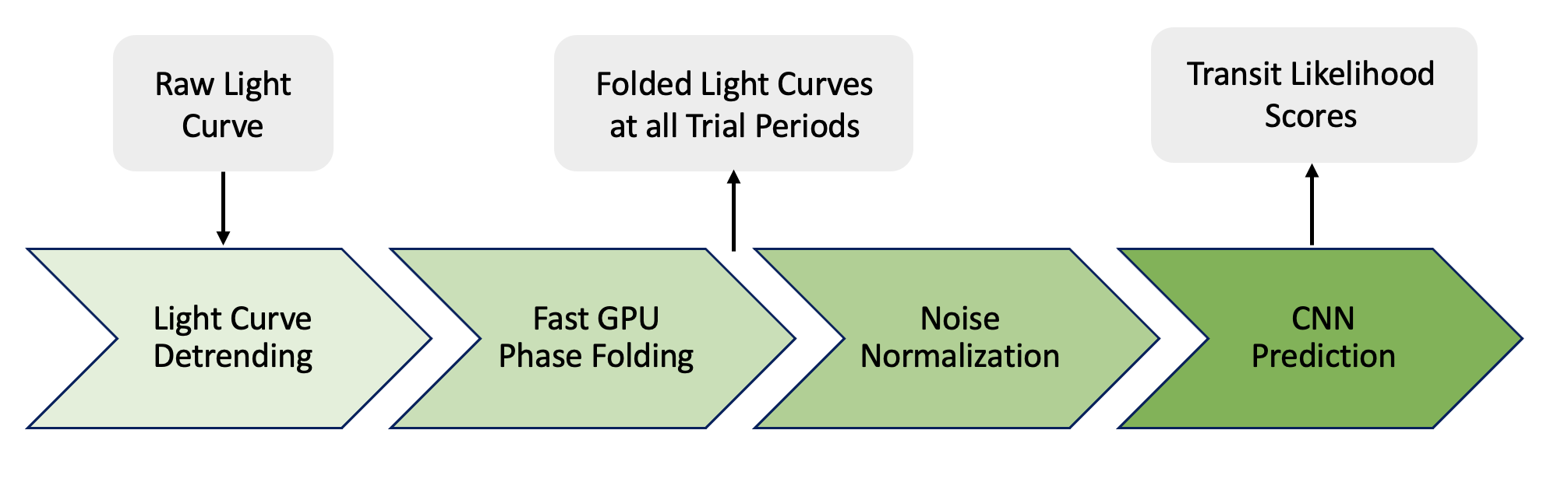}   
    \caption{Fast GPU Phase Folding and CNN (GPFC) Processing Pipeline. The GPFC approach initiates by ingesting a raw light curve and subjects it to detrending. Following this, the light curve is phase folded using a high-precision grid of trial periods. Then the folded results are noise normalized and fed into the CNN, which produces a probability score indicative of the likelihood that the light curve contains a transit event.}
    \label{fig:pipeline-textbox}
\end{figure}

  The architecture of the GPFC method is shown in Fig.~\ref{fig:pipeline-textbox}. The GPFC method comprises of a workflow of pre-processing, GPU parallelized phase folding, noise normalization and CNN transit prediction. To begin with, the GPFC method takes a raw light curve as input, obtained from the Kepler survey in this research, but the method is generic and can be used with other survey data as well. Before analysis, the raw light curve undergoes pre-processing, including detrending and iterative removal of outliers. Next, the light curve is folded at various trial periods, tightly and evenly divided across a designated search range, using the GPU phase folding method. \rt{After phase folding, the GPFC method organizes data samples into bins based on their phase and calculates the average flux for each bin. In a typical 64K data point Kepler light curve, which is evenly distributed into 256 bins, each bin encompasses 256 data points. This approach shows good tolerance to gaps and irregular individual observations in the original Kepler light curve, given that each bin contains a sufficient number of data samples.} The resulting folded \rt{and binned} light curves are normalized to a specific noise level and then passed through a CNN module to evaluate whether any exoplanet transit exists in any of the trial periods.

   As the focus of our research is to search ultra-short-period (USP) exoplanet candidates, we choose to search the period range of less than one day, although the GPFC method can be extrapolated to other period ranges too. 

   A significant challenge of the search for USP planets is the intensive computation needed for detecting signals with short periodicity. Since a priori knowledge of the potential transit period is not available, a vast number of trial periods need to be evaluated. As proposed by \citet{VanderPlas_2018}, to ensure that a period scan does not miss signals in a periodogram, the total number of required sampling for the periodogram with a total observing time window of $T$ can be calculated by Equation \ref{sampling_equation}, 

\begin{equation}\label{sampling_equation}
 N_{sample} = \frac{N_{o}T}{L},
\end{equation}
    where $L$ is the expected width of the signals, and $N_{o}$ is an over-sample factor. This formula can be applied to the \textit{Kepler} survey to determine the sampling precision necessary for detecting USP signals within a \textit{Kepler} light curve. As \textit{Kepler} survey spans an observing window of around 4 years, a typical light curve has total time span $T=4\times 365$ days. An analysis of the 43 confirmed USPs listed in the \textit{Kepler} KOI Catalog reveals that their transit durations lie between $(0.03, 0.09)$ days, prompting us to adopt $L=0.03$. Adhering to Nyquist's Theorem, we set $N_{o} = 2$ to satisfy the minimum sampling criteria. Consequently, to detect \textit{Kepler} USP transits, $N_{sample} = \frac{2 \times 4 \times 365}{0.03} = 97,333$ samples are necessary in the USP period range, which translates to a trial grid precision of as fine as $0.7$ seconds. Maintaining this high precision is essential, as even slight deviations from the transit period, on the order of seconds, can obscure the USP transit signal.

\begin{figure}
  \centering
  \subfloat{\includegraphics[width=0.34\textwidth]{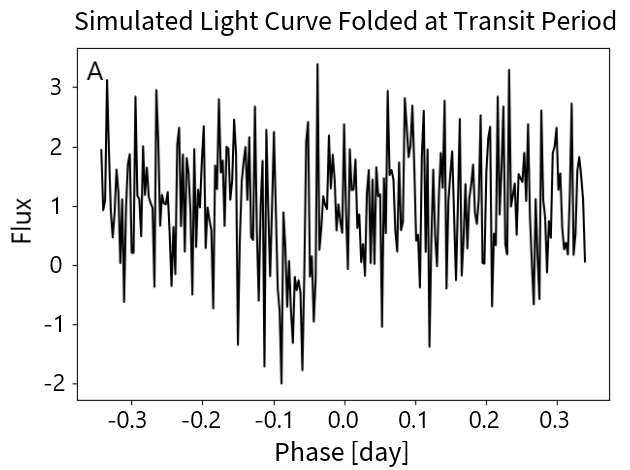}}
  \vfill
  \subfloat{\includegraphics[width=0.36\textwidth]{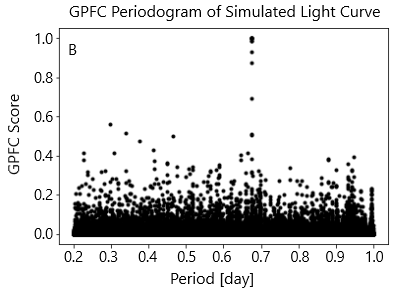}}
  \hfill
  \subfloat{\includegraphics[width=0.36\textwidth]{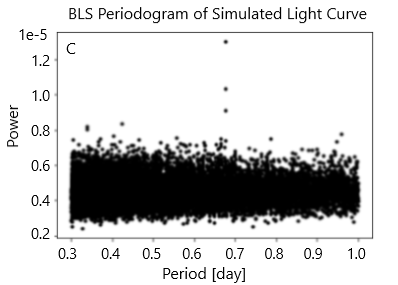}}
  \caption{Example outputs of the GPFC and BLS methods on a simulated light curve. The light curve phase folded at the instrumented transit period is shown in the top panel. The score vs. trial period is illustrated for both GPFC (the middle panel) and BLS (the bottom panel), showing peak scores at the correct period.}
  \label{fig:periodogram-cnn-bls}
\end{figure}

   Based on the above calculation, we configure our phase folding program to uniformly sample 100,000 periods within a representative USP search period defined as $p\in_R [0.2, 1.0]$ days, resulting in a trial period granularity of 0.7 seconds. In tandem, the GPU phase folding algorithm categorizes the light curve to 256 bins, producing 100,000 folded light curves, culminating in $100,000 \times 256$ data points. If a light curve's folding period is the same as the planet's correct transit period, the light curve will manifest a clear transit, as depicted in Fig.~\ref{fig:periodogram-cnn-bls}. Otherwise, the phase misplaced transit signals will become indistinguishable, buried in their surrounding noise after folding.

    Specifically, the noise attenuation achieved through phase folding is  proportional to $\sqrt{N}$, where N represents the total count of the sub-segments folded. This noise attenuation is evident when observing a characteristic \textit{Kepler} light curve with a 4-year observation window. Given an example USP search period $p \in_R [0.2, 1.0]$ days, $N$ can be calculated as: $N=\frac{4 \times 365}{p}$, and $\frac{1}{\sqrt{N}}=\frac{1}{\sqrt{\frac{4 \times 365}{p}}}\in_R [0.01, 0.03]$. This indicates that the noise within the folded light curve diminishes to around $1\%\sim3\%$ of its original magnitude.
  
  Before the folded results are fed into the CNN, we first scale each of the 100,000 folded results to a common noise standard deviation level, as neural networks are often sensitive to the noise level of the input data. \rt{This normalization specifically entails scaling the folded flux values to a standard deviation of 1.0, while maintaining the mean of the folded results. This procedure aligns with the standard deviation of the synthetic data used in training the neural network. }This normalization step standardizes the model prediction scores of the folded results among different targets so that CNN's predictions are comparable across targets.
  
  We then feed the normalized folded results into our CNN, which discerns whether each of the 100,000 folded results contains a transit signal. When such a transit signal exists, we will observe a peak in the model score at the detected orbital period (and we may also see weaker peaks at the harmonics of the orbital period). Because stronger signals result in higher CNN prediction scores, we sort the folded light curves based on the height and width of the peak and choose the best one, which is equivalent to the most probable transit signal among the 100,000 folded results. Fig.~\ref{fig:periodogram-cnn-bls}(A) shows flux values of a simulated folded light curve that has artificial transits at a period of 0.6852 days. As demonstrated in Fig.~\ref{fig:periodogram-cnn-bls}(B) is a graph of CNN score vs. trial period for GPFC. Here, GPFC reports scores at all trails periods, revealing a peak score at the correct transit period. For comparison, Fig.~\ref{fig:periodogram-cnn-bls}(C) illustrates a corresponding graph of power vs. trial period for BLS. This is shown as a reference and will be discussed further in the subsequent Box-fitting Least Squares Periodogram section.

\begin{figure}
  \centering
  
  \begin{minipage}[c]{0.42\textwidth}
    \centering
    \subfloat{\includegraphics[width=\textwidth]{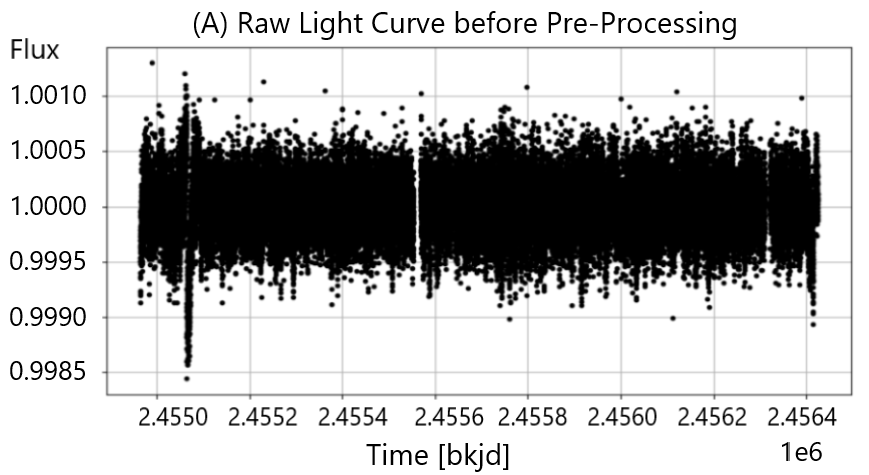}}\label{fig:raw_lc}
  \end{minipage}
  
  \vspace{10pt}
  
  \begin{minipage}[c]{0.43\textwidth}
    \centering
    \subfloat{\includegraphics[width=\textwidth]{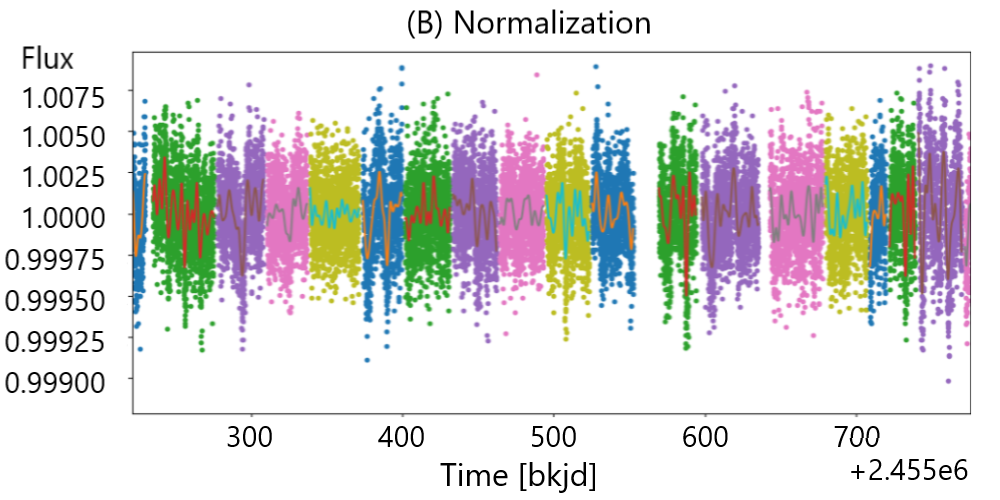}}\label{fig:lc_normalization}
  \end{minipage}

  \vspace{10pt}
  
  \begin{minipage}[c]{0.43\textwidth}
    \centering
    \subfloat{\includegraphics[width=\textwidth]{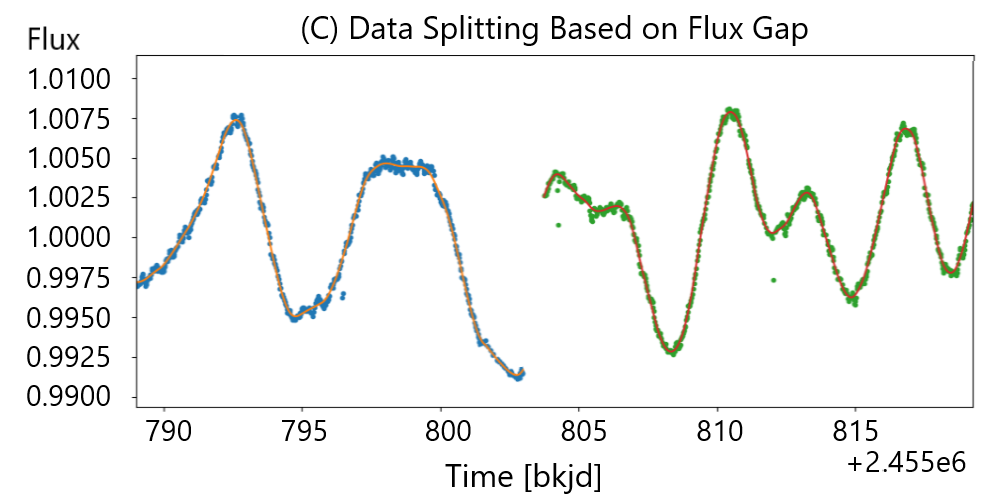}}\label{fig:lc_flux_gap}
  \end{minipage}

  \vspace{10pt}
  
  \begin{minipage}[c]{0.49\textwidth}
    \centering
    \subfloat{\includegraphics[width=\textwidth]{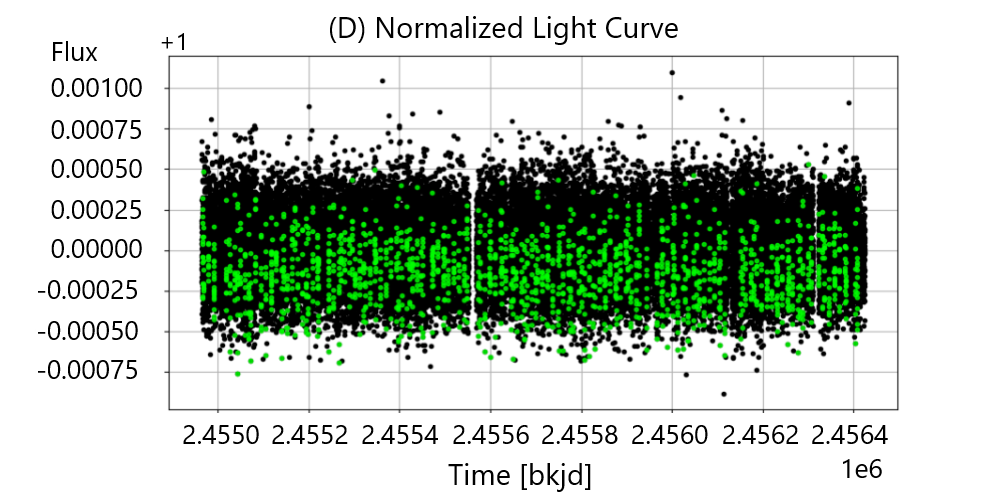}}\label{fig:normed_lc}
  \end{minipage}

  \caption{\textbf{The Data Preprocessing Module:} (A) a raw \textit{Kepler} light curve before preprocessing, (B) preprocessing steps conducted: masking of known transits, segmenting the light curve to multiple data sections, and cropping edges, (C) splitting by flux gap criteria and continuum fitting in each segmented section, (D) the final normalized light curve after preprocessing, where data points in transit windows are plotted in green. }
  \label{fig:normalization}
\end{figure}

\subsection{Pre-processing}

The \textit{Kepler} light curves used in this research are produced by the \textit{Kepler} Science Processing Pipeline \citep{Jenkins_2010}, with each light curve consisting of integrated flux measurements with a cadence of 1,766 second (\textasciitilde29.4 minute) intervals spanning up to four years, in the range of 30,000 to 70,000 epochs. Before they are analyzed by the GPFC system, these \textit{Kepler} light curves are pre-processed in a manner similar to the fitting process illustrated in \citet{Vanderburg_2014}. \rt{First, each light curve is divided into multiple segments based on the time and flux gaps observed within the light curve. As we scan through the light curve, segmentation is triggered if the time interval between two adjacent data points exceeds a predefined time gap threshold, or if the flux value difference between the two adjacent data points exceeds a predefined flux gap threshold. In this study, we set the time gap threshold at 30 times the Kepler data sampling interval (0.020428 days), equivalent to approximately 14.7 hours. This threshold is selected to effectively segment the light curve during relatively long pauses in Kepler observations. Additionally, we set the flux gap threshold at five times the standard deviation of the light curve's flux values. This segmentation strategy is intended to enhance the accuracy of the piecewise fitting process.} For each segment, spline fitting was then performed with all known transits masked to ensure the preservation of transit signals. Throughout this fitting process, outliers exceeding $3\sigma$ were iteratively removed, and spline smoothing parameters were fine-tuned to minimize the Bayesian Information Criterion (BIC). Next, after dividing the light curve by the best-fit spline, we clipped 30 data points from each end of the segment where the spline fits may be skewed. Subsequently, we stitched these segments together, producing a fully detrended light curve. This procedure, illustrated in Fig. \ref{fig:normalization}, effectively detrends the raw light curve and removes low-frequency stellar variability. The pre-processed light curves are subsequently fed into the GPU phase folding module.

\subsection{Fast GPU Phase Folding}

Phase folding increases signal-to-noise ratio of a light curve which maintains periodic signals while reducing non-periodic noises. It is a critical technique for searching small transiting planets which generate weak and shallow transit signals in a noisy background. At the same time, it dramatically reduces the dimensionality of the data making it feasible for further inference with a CNN. To find small ultra-short-period exoplanets potentially overlooked by previous methods, the high precision of the folds is essential because an offset of a few seconds may obfuscate a narrow transit signal. On the other hand, however, folding at a large number of trial periods per light curve is computationally untenable with traditional methods. To attain both folding precision on our trial period grid and practical computational speed, we present the GPU phase folding algorithm which utilizes GPU technology to scale the heavy folding workload with high parallelism and great computational speed. In this research, we used the Nvidia GeForce RTX 3090 Ti, a standard commercial GPU card, to implement our GPU phase folding algorithm. 

A typical phase folding process \citep[e.g.][]{Shallue_2018} consists of three steps: (1) computing modular residuals using the timestamp of each light curve data point modded by a chosen folding period, (2) sorting the flux data points by their modular residuals, and (3) allocating the flux data points equally across a set number of bins. However, this phase folding method doesn't fully leverage GPU parallelism, as the GPU library doesn't support the associated sorting in parallel. On the other hand, since flux data points are binned and averaged immediately post-sorting, the order of the flux data points within each bin is irrelevant and sorting becomes unnecessary.

Based on this analysis, we devised a slightly different phase folding process that can be fully parallelized by GPU. At its core, the objective of phase folding is to group and average flux data points with similar timestamp modular residuals. However, directly mapping of these data points into a pre-determined set of 256 bins is problematic, because certain bins may end up sparsely populated or even entirely empty, while other bins could be densely populated. Such uneven distributions lead to under-populated bins generating noisy average flux values, mainly because these bins become excessively susceptible to anomalies or outliers present in the original flux data points.

Thus we devised a two-tier mapping approach. Our algorithm introduces an intermediary phase with a significantly higher bin count, specifically, 4096 bins. In this phase, the flux data points are mapped to these 4096 bins by the modular residuals of their timestamps. Following this, a merging process is initiated to consolidate these 4096 bins down to the intended 256 bins. The criteria of this merging is to ensure that each of the 256 bins contains an identical number of flux data points. As an example, given a light curve comprising 65,536 data points, each of the 256 bins will accommodate $65536/256=256$ data points. \rt{Note that in our two-tier mapping approach, it's permissible for some of the initial 4096 bins to have a lower number of samples or even be empty. This is because the second-tier mapping effectively consolidates these intermediate bins into the final set of 256 bins, and then calculates the average flux for each.}

 In summary, the initial phase of mapping to the 4096 bins acts as a near-perfect emulation of sorting. Subsequent rebinning to the 256 bins produces homogeneous noise in the binned data to minimize false signals. Noise reduction is further achieved by averaging the flux values within each bin. Note that this rebinning of the 256 bins might introduce minor time step differences on certain occasions, but it does not lead to any missed signal detection. \rt{This was verified using two methods: firstly, through visual inspection of the folded light curves, we confirmed that the outcomes from the two-tier mapping closely resemble those from the traditional sorting-based method. Secondly, by testing known ultra-short-period (USP) planets in the Kepler catalog, we confirmed that our two-tier mapping approach successfully detects transits at the same periods as identified by the traditional method.}
 
 We compared our method with the traditional sorting-based phase folding method and confirmed that our approach exhibits no compromise on accuracy while providing a significantly faster performance by leveraging GPU parallelization.

\begin{figure*}
    \centering
    \includegraphics[scale = 0.65]{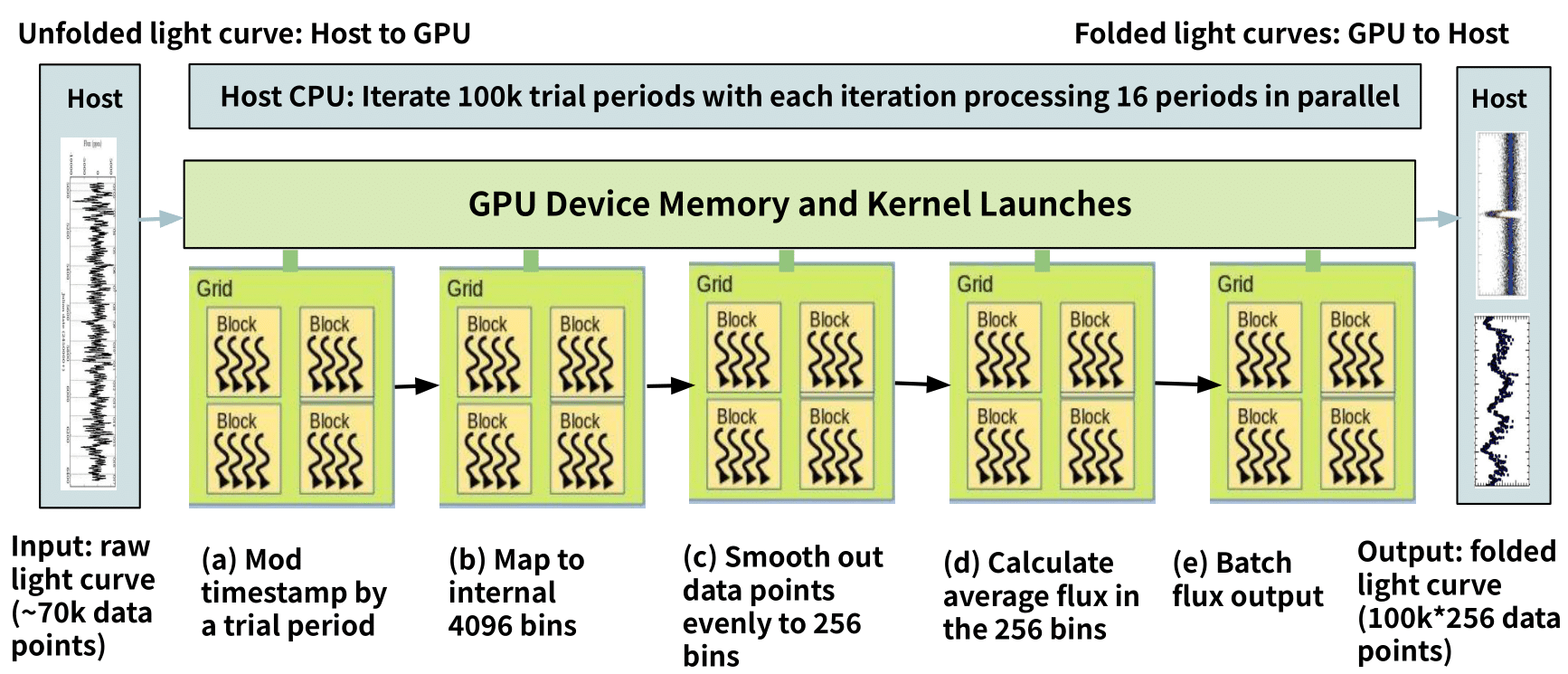}
    \caption{The GPU Phase Folding Module. The GPU phase folding algorithm consists of 5 steps, each of which is optimized through parallel computing. First, (a) the timestamps of the light curve data points are modded by the trial period. Then, (b) the full time span of the trial period is evenly split into 4096 equally-spaced bins, and the data points are mapped into the bins based on the value of their time residuals from (a). Next, (c) the 4096 bins are rebinned into 256 bins, such that each of the 256 bins contains an equal number of data points. The flux values of the data points in each bin are averaged in (d) and filled into its final form, 100,000x256 data points, in (e). The initial mapping to the 4096 bins in (b) serves as a near-perfect emulation of sorting, and the subsequent mapping to the 256 bins and averaging in (c) and (d) create homogeneous noises in the binned data to minimize false signals and achieve noise reduction. The algorithm takes advantage of the GPU's blocks and threads structure such that multiple ($p$) trial periods are folded at the same time. ($p$ was 16 with the GPU we used but it can be more with an advanced GPU with additional memory and thread parallelism.) In other words, for each of the steps from (a) to (e), there are $p$ of that process executing at the same time. } 
    \label{fig:gpu_fold}
\end{figure*}

Thus, as illustrated in Fig. \ref{fig:gpu_fold}, the resulting GPU-optimized phase folding process---which is a mathematical near-equivalent to typical phase folding---is composed of the following five steps. First, we calculate the modular residuals of the timestamp of each light curve data point modded by a given folding period. Secondly, we map the modular residuals to an intermediary 4096 bins. Thirdly, for each of the 4096 bins, calculate the number of flux data points and the sum of the flux values in that bin. Fourthly, scan the 4096 bins from left to right, combine the adjacent bins, and split bins on the boundary as needed, to generate a final layout of 256 bins. This step reallocates the flux data points evenly from the intermediate 4096 bins to 256 bins, with each of the 256 bins containing the same number of flux data points. This step needs to be done sequentially but it's fast because it only needs to convert from 4096 bins to 256 bins instead of converting from 64k light-curve data points to 256 bins. Lastly, we calculate the average flux values for each of the 256 bins, dividing the sum of flux values by the number of data points for each bin.

For each \textit{Kepler} light curve, the GPU phase folding algorithm applies this phase folding process at each of the 100,000 trial periods. The algorithm divides the 100,000 trial periods into batches of 16, processing each batch in parallel until all periods are completed. With the Nvidia GeForce RTX 3090 Ti card that is used in our research, we let the GPU kernel execute 16 periods simultaneously, which uses close to the entire 11 GB GPU device memory on the card. The final output of the GPU phase folding algorithm is a data array with dimensions $100,000\times 256$, which represents the folded result at each of the $100,000$ evenly-spaced trial periods. The GPU Phase folding algorithm takes \textasciitilde5 seconds to finish 100,000 folds for a typical \textit{Kepler} light curve.

\subsection{GPU Phase Folding Performance Tuning}

Because the performance of the GPU algorithm is key to our GPFC system, we exploited various techniques to maximize parallelism and optimize its performance. Although we have briefly touched on some of the performance tuning, below is a description of the techniques used.

\begin{enumerate}
  \item \textbf{Maximization of GPU kernel’s parallelism in the phase folding algorithm.}
 Our fast GPU phase folding algorithm was composed of a series of GPU kernel function launches. Nvidia GeForce RTX 3090 Ti GPU card supports a maximum of $65535\times 65535\times 65535$ block-n-threads parallelism. We used the first dimension to parallelize across the 100,000 periods, with 16 phase folds conducted simultaneously. For each of the functions of which the GPU phase folding algorithm consists, we assigned the other two dimensions on the raw light curve data points such that they are all processed by the GPU kernel simultaneously with its thread pool.

  \item \textbf{Minimization of memory transfer overhead between CPU and GPU.}
Although GPU-based executions are fast, the data transfer between CPU and GPU is often an expensive step. To reduce memory transfer overhead, we employed a zero-memory-copy technique, and batched small memory transfers to maximize memory bandwidth utilization. We also minimized the GPU memory read-write time by maximizing the amount of parallel memory accesses. These memory optimizations reduced memory transfer time from 10 seconds to 0.25 seconds in our algorithm.

  \item \textbf{Utilization of atomic read-write operations to speed up thread serialization.}
With significant concurrent execution of numerous GPU kernel threads, it is imperative to implement thread serialization when altering data stored at a shared memory location. To address this requirement, we leveraged the inherent atomic operation hardware support of the NVIDIA GeForce RTX 3090 Ti card as our chosen synchronization tool. Notably, the utilization of atomic operations resulted in a twelvefold speed increase in comparison to the conventional locking mechanism.

\end{enumerate}

Table~\ref{tab:GPU_perf_profiling} shows a breakdown of the final runtime of our GPU phase folding algorithm. As seen in this example, the program spent most of the time on memory read/write operations within the GPU device (3.66 seconds), then on calculations within the GPU device (0.66 seconds), followed by the batch memory transfer from the GPU card to the CPU host (0.25 seconds), then by memory allocation and free operations, mostly spent on the GPU device (0.18s). \rt{Lastly, initializing the program on the host CPU took a small time slice of 0.042 seconds. The total execution time amounted to 5.00 seconds. Note that this measured duration pertains to the folding and binning of a light curve consisting of 70,000 epochs. The execution time is subject to slight variations depending on the length of the Kepler light curves, with a marginal reduction in duration for shorter light curves. Roughly speaking, the maximum running time for processing Kepler light curves is around 5 seconds.}

\begin{table}
\begin{tabular}{ p{6cm} p{1.3cm}  }
 \hline
 \hline
\multicolumn{2}{ c }{GPU Phase Folding Performance Profiling} \\
 \hline
 \hline
 Runtime Breakdown & Time (sec)\\
 \hline
Memory reads and writes on GPU device & 3.66\\
Running GPU Kernel functions & 0.66\\
Memory transfer from GPU to host & 0.25\\
 \hline
Total Memory allocation/free (host and GPU)& 0.18\\
\hspace{4mm}Memory allocation on GPU (heap) &  0.17\\
\hspace{4mm}Memory allocation on host (stack) &  0.0097\\
\hspace{4mm}Memory free on GPU (heap) &  0.00058\\
 \hline
Read input data files on host & 0.11\\
Write output to disk on host & 0.098\\
 \hline
\rt{Initialization on host CPU} & \rt{0.042}\\
 \hline
Total time & 5.00\\
 \hline
\end{tabular}
\caption{GPU phase folding performance profiling breakdown. In this example, the longest time of the GPU program was spent on memory read and write operations within the GPU device (3.66s). The second longest time was spent on calculations in GPU kernel functions (0.66s), followed by the batch memory transfer from the GPU device to the CPU host (0.25s), then by memory allocation and free operations, mostly spent on the GPU device (0.18s). \rt{Additionally, initializing the program on the host CPU took a small time slice of 0.042 seconds.} The total execution time amounted to 5.00 seconds.}
\label{tab:GPU_perf_profiling}
\end{table}

\subsection{Deep Neural Network}

After the 100,000 folds of a light curve are noise-normalized, as described in the introduction section, they are ready to be fed into the convolutional neural network to discern whether they contain a transit signal at any of the trial periods.

\begin{figure}
    \centering
    \includegraphics[scale = 0.45]{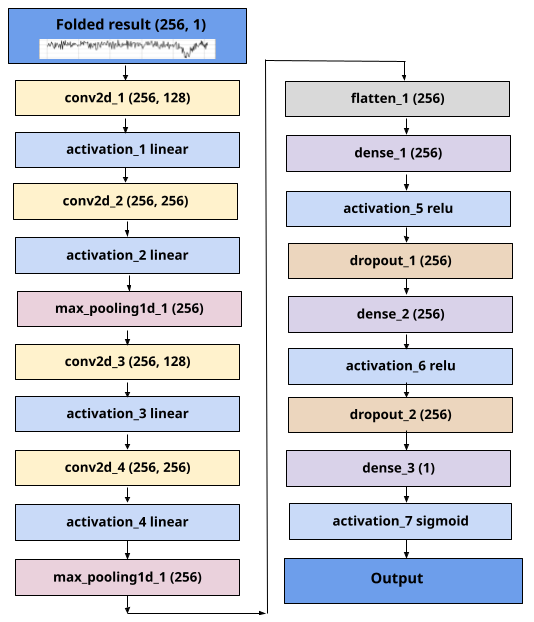}
    \caption{The CNN Module. As input, the CNN takes a single noise normalized 256-length light curve fold and the CNN outputs a confidence score that the folded input contains a transit signal.}
    \label{fig:cnn_model}
\end{figure}
The design of our convolutional neural network architecture consists of 19 total layers as illustrated in Fig.~\ref{fig:cnn_model}. To start with, the initial 1-D input data undergoes reshaping into a 2-D tensor, preparing it for subsequent 2-D convolutional operations. Taking into account the periodic nature of the data, we design a model incorporating a circular convolution layer, a convolution with edge wrapping sourced by \citet{schubert19}. The initial circular convolution layer employs 128 filters, followed by another with 256 filters. Subsequently, the 2D tensor is reverted to its original 1D form, and a global-max-pooling layer is introduced to retain the most significant values. Following this, the same sequence of operations is repeated, this time utilizing larger kernel sizes to capture broader spatial patterns within the data. Afterwards, the data is flattened and passed through fully connected layers with the ReLU activation function. Dropout layers are incorporated to mitigate overfitting during the training process. Ultimately, a dense layer with the sigmoid activation function furnishes the model's prediction score. The output of the model is a probability score representing the likelihood that the inputted data contains a planetary transit signal. We construct our model on the top of the open source TensorFlow library \citep{Abadi_TensorFlow_2016}, and we use the Adam optimization algorithm \citep{Kingma_Adam_2014} to minimize the cross-entropy error function. We train the neural network with a learning rate of $\mathrm{10}^{-6}$ and a batch size of 32 across 90 epochs. \rt{We train the neural network with a learning rate of $\mathrm{10}^{-6}$ and a batch size of 32 across 90 epochs. The final parameters were selected based on approaches used in similar applications, along with experimentation, validation, and testing. For future work, hyperparameter tuning methods such as grid search or random search can be employed to systematically explore further optimizations}.

\subsection{Light Curve Simulation}

The signal-to-noise ratio (SNR) of a planetary transit detection in a given light curve can, in the simplest case, be approximated by Formula \ref{snr_equation} \citep{von_Braun_2007},

\begin{equation}\label{snr_equation}
 \alpha = \frac{d}{\sigma}\sqrt{n\frac{L}{p}}
\end{equation}

where $d$ is the transit depth, $\sigma$ represents the photometric measurement uncertainty in relative flux per data point, assuming it is the same for all data points. $p$ is the transit period, and $L$ is the transit duration. $n$ equals the total number of data points in observation, therefore, equivalently, $n\frac{L}{p}$ equals the number of data points observed during transits. The assumption of this equation is that there exists only white noise and no statistically correlated (red) noise.

\begin{figure}
  \centering
  \subfloat{\includegraphics[width=0.23\textwidth]{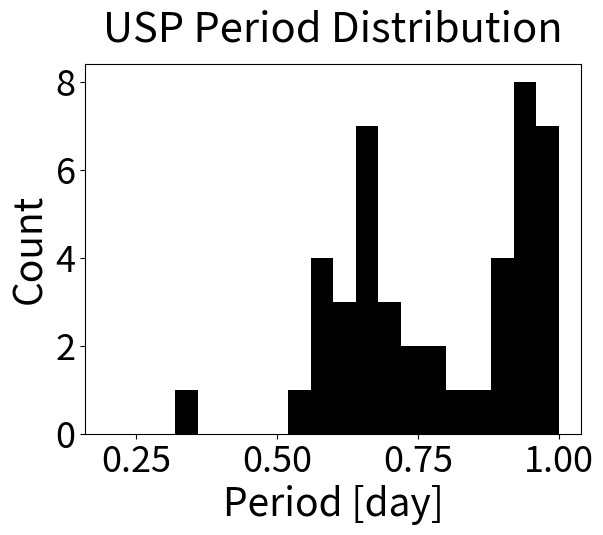}}\label{fig:usp_per_dist}
  \hfill
  \subfloat{\includegraphics[width=0.235\textwidth]{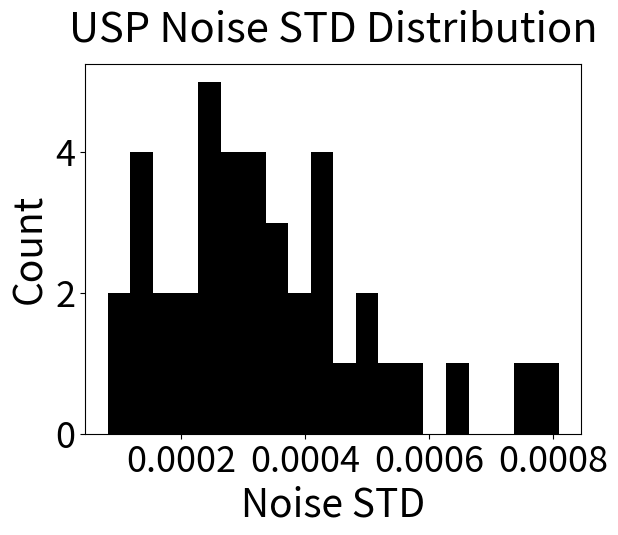}}\label{fig:usp_std_dist}
  \vfill
  \subfloat{\includegraphics[width=0.23\textwidth]{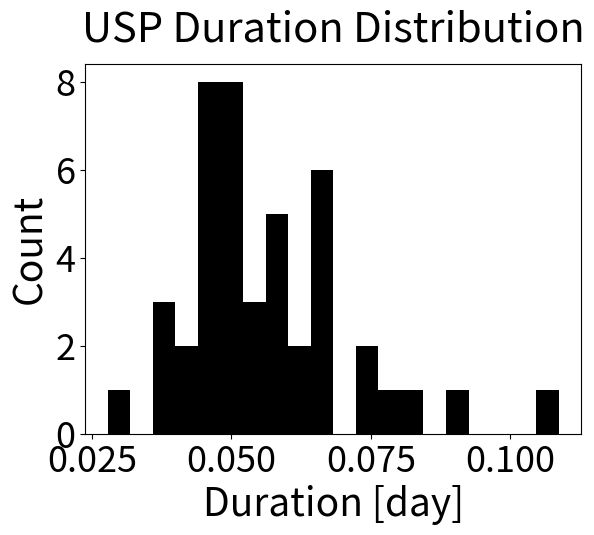}}\label{fig:usp_dur_dist}  \hfill
  \subfloat{\includegraphics[width=0.23\textwidth]{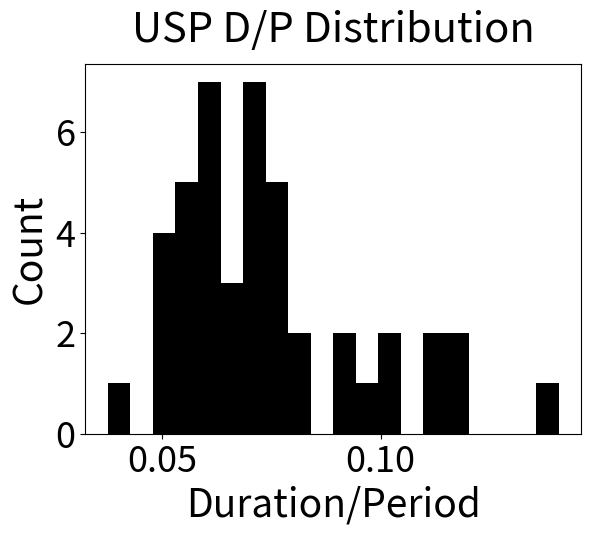}}\label{fig:usp_durOverPer_dist}
  \caption{The parameter distributions for the 43 confirmed USPs from the Kepler Objects of Interest (KOI) Catalog. (a) Period, (b) Noise, (c) Duration, and (d) Duration/Period. These parameters are used as a guide for our simulation of light curves.}
  \label{fig:usp_param_dist}
\end{figure}

To create simulated light curves as realistic as possible, we studied the statistics of the parameters of real \textit{Kepler} light curves and used them as guidance for the simulation. Since our objective is to simulate light curves generated by USP exoplanets, we based our simulation on the parameter distributions observed in the 43 confirmed USPs listed in the Kepler Objects of Interest (KOIs) in the Kepler Input Catalog (KIC), as depicted in Fig.~\ref{fig:usp_param_dist}.

Specifically, we measured the following parameters: orbital period, transit duration, transit duration over orbital period, and the relative flux 1-$\sigma$ uncertainty in the detrended light curves. From the confirmed USPs, we observe that the parameter distributions are as follows: orbital period $p \in_R (0.3, 1.0)$ days, standard deviation of the noise $\sigma \in_R (0.0001, 0.0005)$, transit duration $L \in_R (0.03, 0.09)$ days, and transit duration over orbital period $\frac{L}{p} \in_R (0.04, 0.12)$. These distributions are listed in Table~\ref{tab:usp_param_distribution}. 

\begin{table}
\begin{tabular}{ p{4.4cm} p{3.1cm}  }
 \hline
 \hline
\multicolumn{2}{ c }{USP Transit Parameter Distribution in Kepler} \\
 \hline
 \hline
 Orbital period (days) & $p \in_R (0.3, 1.0)$\\
 Transit duration (days) & $L \in_R (0.03, 0.09)$\\
 Transit duration to period ratio & $\frac{L}{p} \in_R (0.04, 0.12)$\\
 Standard deviation of noise & $\sigma \in_R (0.0001, 0.0005)$\\
 \hline
\end{tabular}
\caption{USP parameter distribution gathered from the 43 confirmed USPs in the \textit{Kepler} KOIs.}
\label{tab:usp_param_distribution}
\end{table}

\begin{figure}
    \centering
    \includegraphics[scale = 0.47]{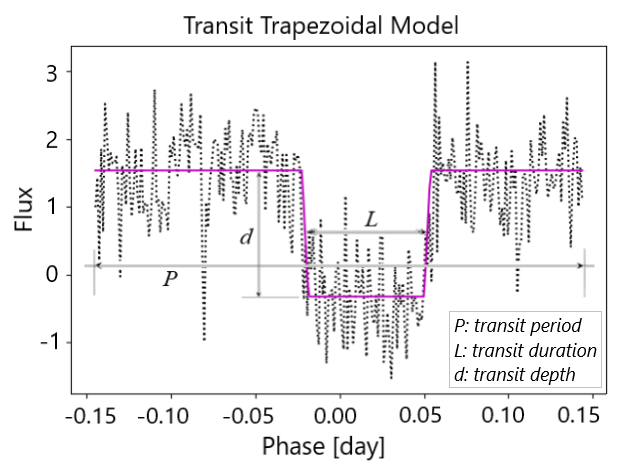}
    \caption{Trapezoidal Model.  A trapezoidal shape is incorporated into Gaussian noise to simulate planetary transit signals within a light curve. The input parameters for the trapezoid are determined based on the actual USP parameter distributions extracted from \textit{Kepler} KOIs. $P$ and $L$ denote the orbital period and the transit duration, respectively. The transit depth, denoted as $d$, is computed given a specific SNR, while the ratio between the trapezoid's short base and long base is randomly chosen $r \in_R (0, 1.0)$.}
    \label{fig:trapezoid_model}
\end{figure}

 We simulate a transit signal with a trapezoidal model as shown in Fig.~\ref{fig:trapezoid_model}. The trapezoidal parameters are determined based on the real USP parameter distributions gathered above. The trapezoid ratio between the short base and the long base is a random number $r \in_R (0, 1.0)$. The transit duration is measured at the half depth of the trapezoid. Given a simulation SNR target value $\alpha$, the transit signal depth $d$ can be derived from Formula \ref{snr_equation}.

\subsection{Synthetic Data Set for Neural Network Training}

As the 43 confirmed USPs in the \textit{Kepler} survey do not constitute a sufficiently large dataset for effective CNN training, we created an extensive training dataset, comprising two million synthetic light curves divided into one million \rt{transit} and one million \rt{non-transit} samples. Each sample set consists of vectors with a length of 256, simulating folded light curves that the CNN is designed to process.

In a foundational model, Gaussian noise can be employed to simulate \rt{non-transit} samples, while the injection of a transit signal into Gaussian noise yields \rt{transit} samples. However, in practice, Gaussian fluctuations can occur especially when light curve segments are repetitively stacked or folded at a specific period. These fluctuations can sometimes resemble transits with low SNRs. Therefore, in addition to pure Gaussian noise, we intentionally introduce \rt{non-transit} samples by injecting transits with low SNRs. This method optimizes the CNN's capability to differentiate between significant Gaussian fluctuations and genuine transits. Empirically, we chose a cutoff SNR of 5; training data produced with a corresponding SNR greater than 5 are labeled as positive, while those with an SNR below 5 are labeled as negative. We also confirmed that our CNN trained with an SNR cutoff of 5 misclassifies Gaussian noise as positive signals in fewer than 0.001 percent of instances.

\begin{figure}
  \centering
  
  \begin{minipage}[c]{0.4\textwidth}
    \centering
    \subfloat{\includegraphics[width=\textwidth]{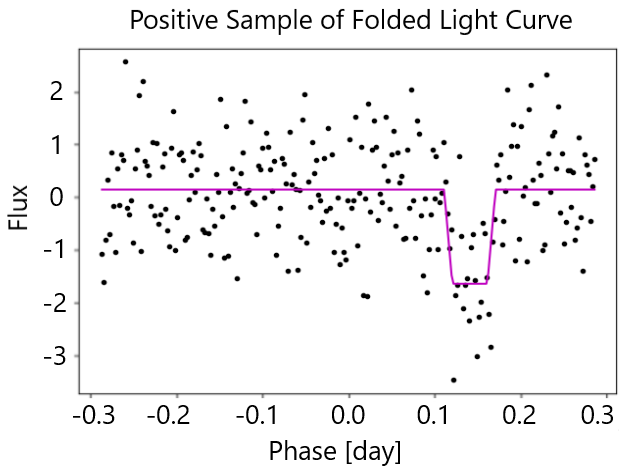}}
  \end{minipage}

  \vspace{10pt}
  
  \begin{minipage}[c]{0.4\textwidth}
    \centering
    \subfloat{\includegraphics[width=\textwidth]{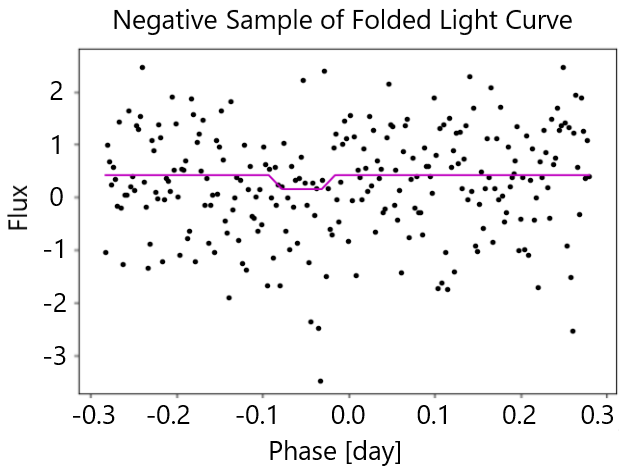}}
  \end{minipage}

  \caption{Simulated folded light curves for CNN training. The upper plot demonstrates a \rt{transit} sample generated by injecting a trapezoidal model into Gaussian noise with period $p=0.569975$, duration $L= 0.0593157$, and SNR $\alpha=9.2$. The lower plot demonstrates a \rt{non-transit} sample generated by injecting a low SNR trapezoidal signal to Gaussian noise with period $p=0.386515$, duration $L= 0.0528367$, and SNR $\alpha=1.6$.}
  \label{fig:simlc-folded}
\end{figure}

To create \rt{transit} samples, we inject a trapezoidal model into randomly generated Gaussian noise of length 256. And we generate \rt{transit} samples in the range of SNR $\alpha \in_R [6, 10]$. Fig.~\ref{fig:simlc-folded} demonstrates the generated \rt{transit and non-transit} samples, respectively.

\rt{Our synthetic dataset represents a simplified model of the complex noises found in actual light curves, which include correlated noise from various sources as well as Gaussian noise. This dataset is confined to scenarios involving planet transits and non-transits, omitting signals characteristic of other common false positive situations. However, considering that the primary goal of this simulation test is to assess the timing performance of our method, this simplified model is adequate for our purposes. We used the same dataset for BLS comparison, ensuring fairness in evaluating relative differences. Nevertheless, it's worth noting that the absolute performance indicated by this model is likely more optimistic than what would be observed with real light curve scenarios.}

\begin{figure}
  \centering
  
  \begin{minipage}[c]{0.41\textwidth}
    \centering
    \subfloat{\includegraphics[width=\textwidth]{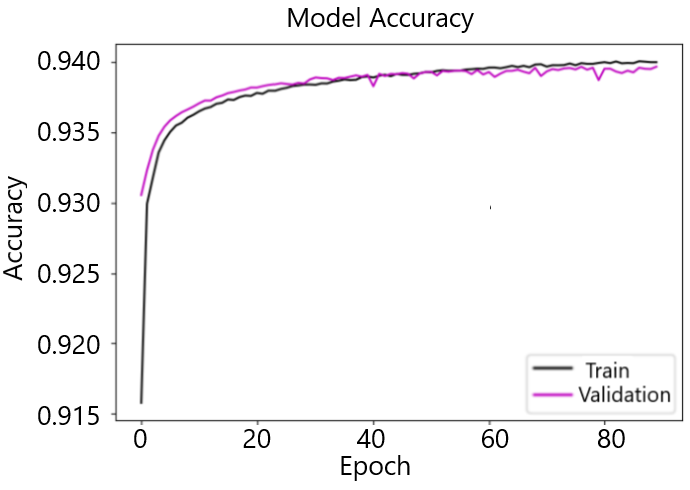}}\label{fig:training_curve_accuracy.png}
  \end{minipage}
  
  \vspace{10pt}
  
  \begin{minipage}[c]{0.41\textwidth}
    \centering
    \subfloat{\includegraphics[width=\textwidth]{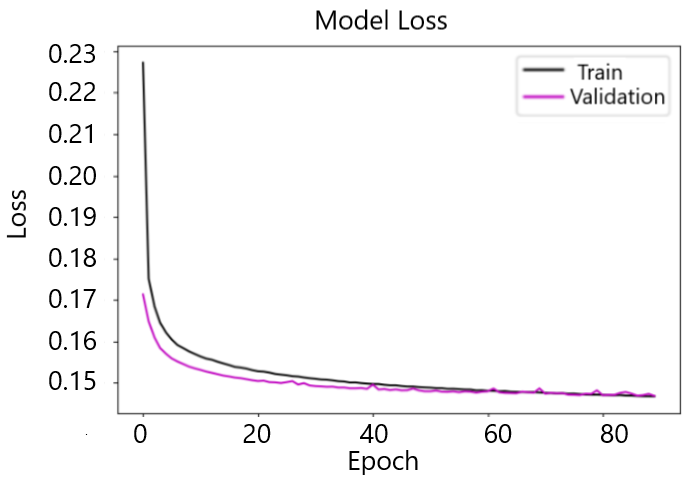}}\label{fig:training_curve_loss}
  \end{minipage}

  \caption{The training curves (accuracy and loss) for the CNN model on the training and validation sets, respectively, across 90 epochs.}
  \label{fig:training_curve}
\end{figure}

By following this process, we generate a dataset comprising one million \rt{transit} and one million \rt{non-transit} samples, which are subsequently randomly partitioned into training (80\%), validation (10\%), and testing (10\%) sets. Following the neural network training, our best model achieves an accuracy of 94.0\%. For a detailed view of the training progress, please refer to Fig.~\ref{fig:training_curve}, which presents the complete training curve across the 90 epochs.

\section{Comparison}

In this section, we conduct a comparative analysis between the GPFC method and the BLS method, using simulated unfolded \textit{Kepler} light curves to evaluate their characteristics and performance metrics. 

To closely replicate real \textit{Kepler} light curves, all the light curves we generated have a length of n = 65536 epochs and a time interval of 0.0204 days (\textasciitilde29.4 minutes) between consecutive data points.

We use Gaussian noise to simulate light curves that do not contain transits. Conversely, for simulating planetary transits within the light curve, we inject a series of trapezoidal shapes into Gaussian noise. These trapezoids are parameterized based on the provided values for orbital period, transit duration, transit epoch, and SNR.

\subsection{The Box-fitting Least Squares Periodogram}

To make a comparison with our GPFC method, we employ the widely recognized Box-fitting Least Squares (BLS) algorithm developed by \citet{Kov_cs_2002}. BLS is a periodogram that phase folds a light curve and fits a transit-like rectangular box, aiming to minimize the fitting squared error. BLS searches for the transit signals on the frequency space, and provides a periodogram power value for each trial frequency. To evaluate the likelihood of a transit-like signal existing in a particular light curve, a Signal Detection Efficiency (SDE) score is computed, as illustrated in the following equation \citep{Kov_cs_2002}:

\begin{equation}\label{SDE_equation}
SDE = \frac{P(f_{max}) - \langle P \rangle}{sd\langle P \rangle}
\end{equation}
where $P$ is the Box-fitting Least Squares periodogram function, $f_{max}$ is the frequency at which the highest power occurs, $\langle P \rangle$ is the mean of the periodogram powers, and $sd\langle P \rangle$ is the standard deviation of the periodogram powers. 

In short, given all the periodogram powers across the frequency space, the $SDE$ value represents the number of standard deviations above the mean of the maximum power. The BLS method also takes parameters including the number of bins, the frequency range, and the range of the fractional duration over the period. In our testing, we set the number of bins used by BLS to 256. The specifications for the frequency range and the range of fractional duration over the period are derived from the distributions illustrated in Fig.~\ref{fig:usp_param_dist}, adopting the same criteria as used for the GPFC method.

\subsection{Accuracy of GPFC vs. BLS}
\begin{figure}
    \centering
        \includegraphics[scale =0.24]{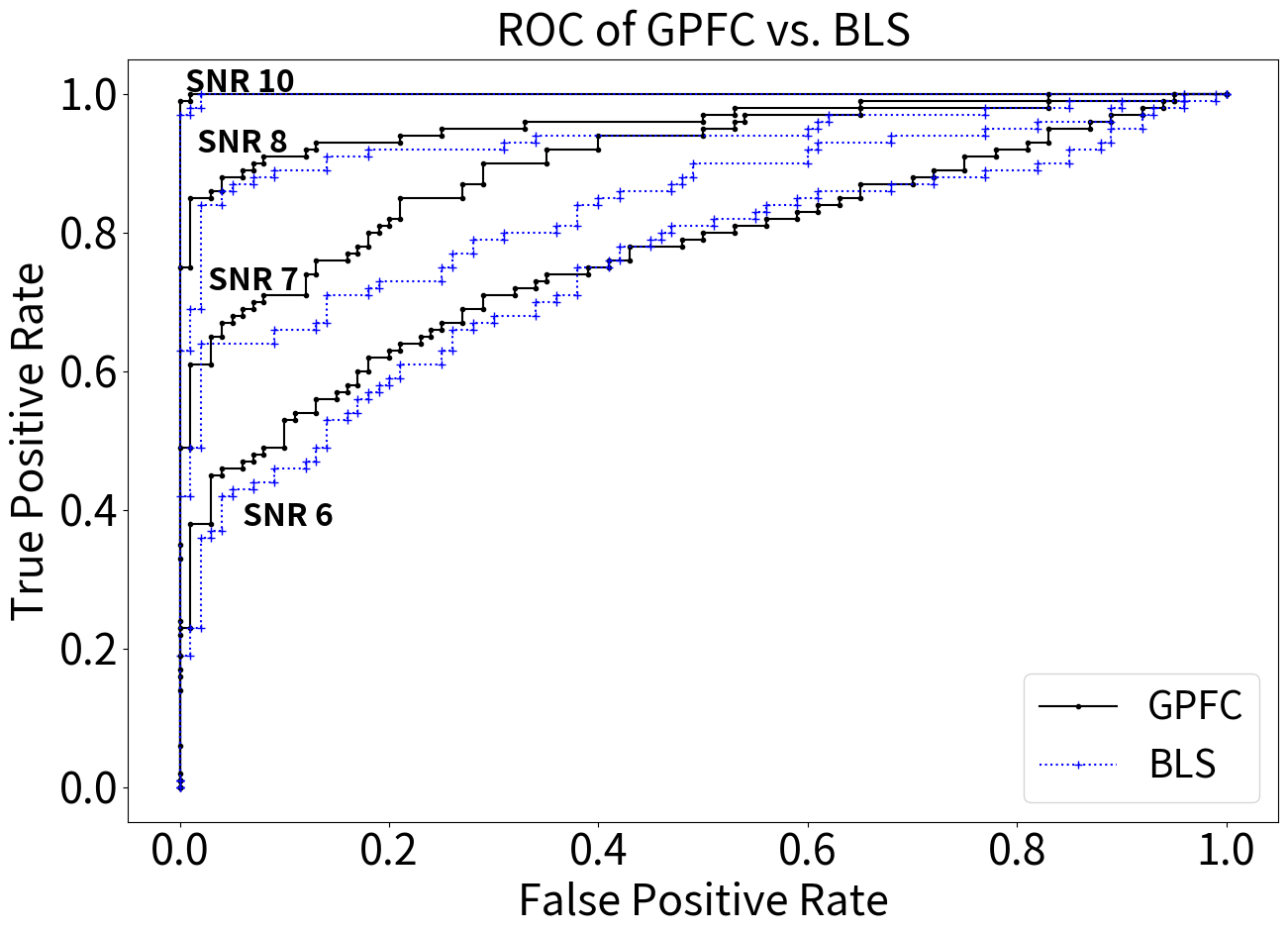}
    \caption{The Receiving Operating Characteristic (ROC) curves for GPFC and BLS, evaluated at 100,000 frequencies, span SNRs of 6, 7, 8, and 10. As the threshold for classifying \rt{transit and non-transit} outputs varies, there is a trade-off between the true positive rate and the false positive rate, shown by the ROC curve. The GPFC method shows stronger ability to distinguish \rt{transit and non-transit} light curves for SNR below 10.}
    \label{fig:roc-bls-vs-cnn}
\end{figure}

\begin{figure}
    \centering
    \includegraphics[scale = 0.49]{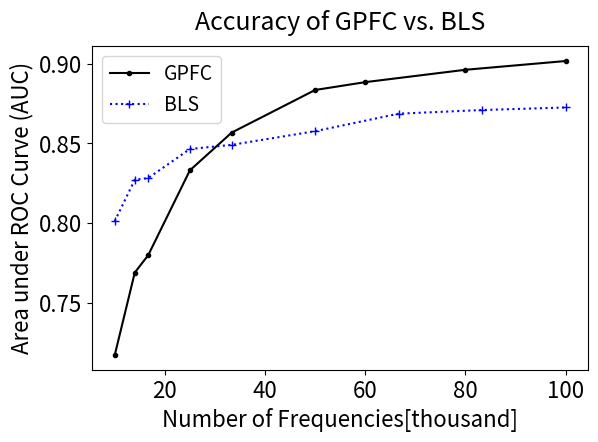}
    \caption{The accuracy of GPFC and BLS as the number of trial frequencies vary. Here, accuracy refers to the ability to distinguish \rt{transit and non-transit} light curves, as represented by the area under the ROC curve.}
    \label{fig:accuracy-auc-cnn-vs-bls}
\end{figure}

\begin{figure}
    \centering
    \includegraphics[scale =0.24]{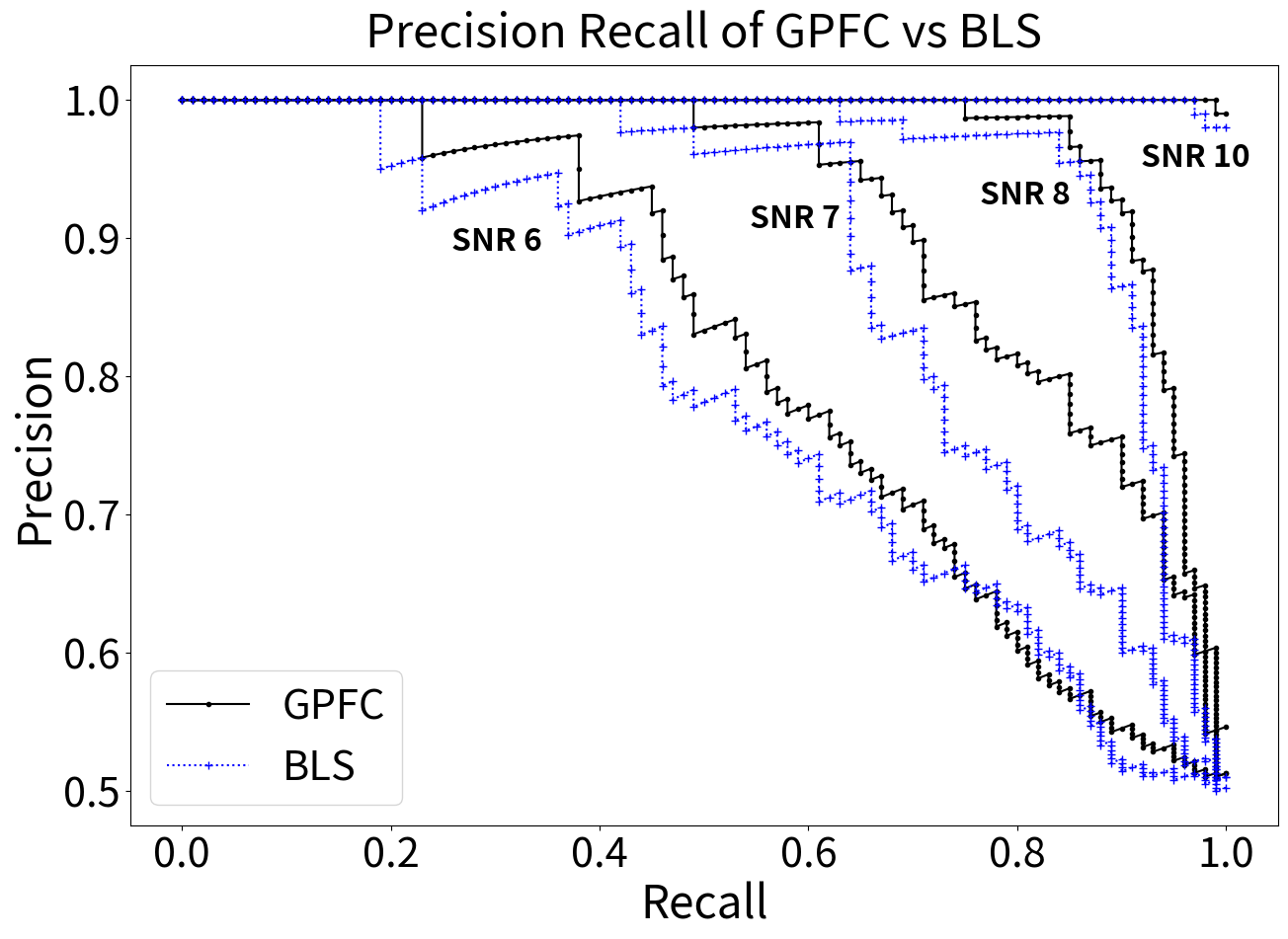}
    
    \caption{The Precision-Recall curves for GPFC and BLS, evaluated at 100,000 frequencies over SNRs 6, 7, 8, and 10 are shown. The GPFC method achieves higher precision than BLS at the same recall in the curves with SNR below 10.}
    \label{fig:precision-recall}
\end{figure}

Next, we analyze the capability of the two methods to detect exoplanet signals in light curves. We posit that the underlying mechanisms that GPFC employs makes it a more powerful detection system because (1) our model evaluates light curves with a deep neural network, which, if trained properly, has the capability for stronger spatial awareness in recognizing transit-like signals than a least-squares approach, and (2) our neural network is trained on a trapezoidal model with white noise, which is a closer representation of a true transit shape than a box-car function. 

To assess the efficacy of the GPFC approach relative to the BLS method, we evaluate their capabilities to distinguish between true and false signals that we simulated across a range of SNRs. For BLS, this capability is reflected by the distinction in the $SDE$ scores assigned to \rt{transit vs. non-transit} light curves; for GPFC, it is represented by the disparity in maximum model scores between \rt{transit and non-transit} light curves. For each method, by selecting various thresholds to demarcate  \rt{transit from non-transit} predictions, we can construct a Receiving Operating Characteristic (ROC) curve. The more powerful a given classifier is, the more its ROC curve will be higher and toward the left (Fig. \ref{fig:roc-bls-vs-cnn}), denoting a larger true positive rate and lower false positive rate across different cutoffs. Thus, we can use the area under the curve (AUC) of the ROC as an overall representation of how accurately a model distinguishes true and false signals. Fig.~\ref{fig:accuracy-auc-cnn-vs-bls} illustrates that the AUC of GPFC, which is employed with 100,000 trial periods, outperforms that of BLS, demonstrating that GPFC is a stronger general classifier of transit signals than BLS. The precision-recall results in Fig.~\ref{fig:precision-recall} similarly show that GPFC exhibits higher performance ability in the tradeoff between true positive rate and false positive rate.

\begin{figure}
  \centering
  
  \begin{minipage}[c]{0.405\textwidth}
    \centering
    \subfloat{\includegraphics[width=\textwidth]{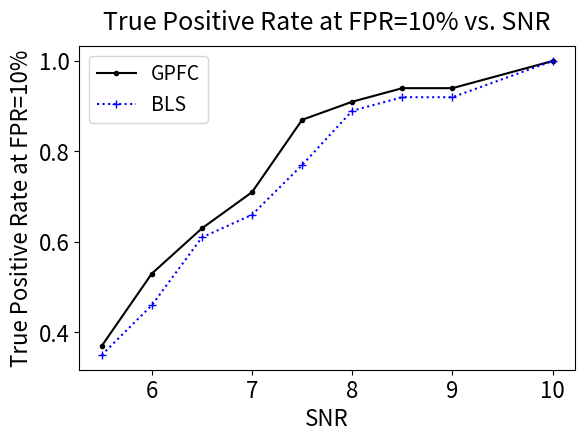}}\label{fig:accuracy-vs-snr-completeness}
  \end{minipage}
  
  \vfill
  
  \begin{minipage}[c]{0.4\textwidth}
    \centering
    \subfloat{\includegraphics[width=\textwidth]{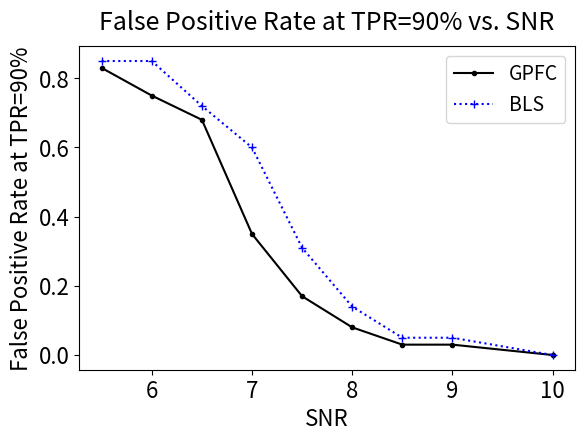}}\label{fig:accuracy-vs-snr-fpr}
  \end{minipage}

  \caption{Performance comparison of GPFC and BLS in two scenarios. (a) at the same false positive rate of 10\%,  the recall is higher for GFPC than BLS for SNR between 5.5 to 10. (b) at the same high recall such as 90\%, the false positive rate is lower for GFPC than BLS for SNR between 5.5 to 10. GPFC outperforms BLS in terms of both metrics: true positive rate when false positive rate is at 10\%, and false positive rate at true positive rate of 90\%. }
  \label{fig:accuracy_over_snr_tpr_fpr}
\end{figure}

While AUC represents overall ability to distinguish across the true positive rate vs. false positive rate tradeoff, in practical research it is often the case that either the true positive rate or the false positive rate is given prioritized importance as a result corresponding to the specific use-case. Therefore, we illustrate a second comparison between GPFC and BLS in Fig.~\ref{fig:accuracy_over_snr_tpr_fpr}, splitting off these two types of situations. When the false positive rate is prioritized (10\% FPR), the resulting true positive rate is shown in the top panel of Fig.~\ref{fig:accuracy_over_snr_tpr_fpr}, whereas when the true positive rate is prioritized (90\% TPR), the resulting false positive rate is shown in the bottom panel of Fig.~\ref{fig:accuracy_over_snr_tpr_fpr}. GPFC shows a distinct advantage over BLS in both metrics: the true positive rate at 10\% false positive rate, and the false positive rate at 90\% true positive rate.

\subsection{Speed of GPFC vs. BLS}

\begin{figure}
    \centering
    \includegraphics[scale = 0.48]{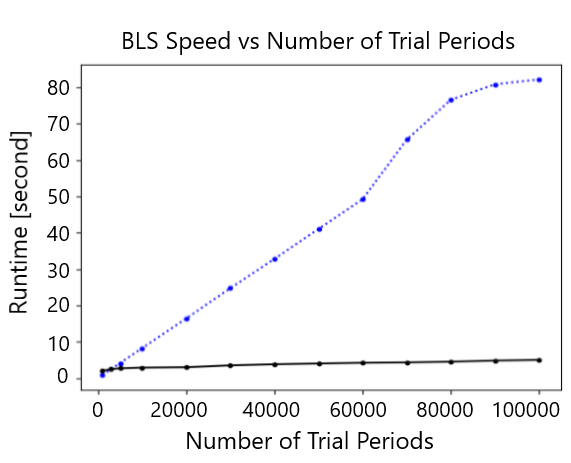}
    \caption{The speed of BLS vs. the number of trial periods searched in the USP period range of $[0.2, 1]$ days. The blue dotted line shows a roughly linear relationship between the runtime of BLS for one light curve as the number of trial periods searched varies. The fast CYTHON-accelerated BLS implementation provided by AstroPy is used for runtime comparison, which takes 80 seconds of runtime at 100,000 trial periods. On the other hand, the GPFC method processes one light curve in 6 seconds at 100,000 trial periods, which is 15 times faster.}
    \label{fig:bls-speed-upto-100k}
\end{figure}

The GPU phase folding algorithm takes \textasciitilde5 seconds per light curve with length 65536, and the CNN takes \textasciitilde6 seconds to evaluate the 100,000 folded results. Thus, to process one light curve with the GPFC method takes \textasciitilde11 seconds. Note that when using the GPFC method to process a large number of light curves, the speed is reduced to 6 seconds per light curve because the CNN and GPU Phase Folding algorithm can be run simultaneously---as the GPU phase folds a light curve, the CNN can process the previously folded light curves in parallel. With the 6 seconds per light curve speed, the entire catalog of the 150,000 \textit{Kepler} main-sequence stars can be searched in just over 10 days. On the other hand, as \citet{Kov_cs_2002} mentions, the performance of the BLS periodogram varies based on the number of frequencies scanned in its frequency space. The run-time for BLS with the same search accuracy on the same light-curves would be three orders of magnitude higher. To speed up the runtime comparison, we further assessed a fast BLS implementation provided by AstroPy (docs.astropy.org), which is accelerated by CYTHON. Fig.~\ref{fig:bls-speed-upto-100k} shows the AstroPy BLS method to have a run-time of \textasciitilde 80 seconds at frequency of 100,000, which is 15 times slower than GPFC.

The performance advantage of GPFC over BLS is attributed to a few factors built in the design of the GPFC system. To search for a flux "dip" in a light curve, the BLS algorithm constructs a three-level nested loop to exploit an exhaustive scan over all possible transit start positions, transit durations, and all trial frequencies. By using a CNN to vet light curves, we eliminate the two loops of exhaustive search over transit start positions and transit durations, leveraging CNN's advantage to recognize objects that contain spatial structure. For the loop over trial periods, the GPU-based folding algorithm, the GPU's capability for parallelism allows it to phase-fold multiple frequencies at the same time, greatly boosting performance.

\section{GPFC Applied to Real \textit{Kepler} Light Curves}

\subsection{Kepler Light Curve Processing}

Next, we confirm that GPFC works on real \textit{Kepler} light curves. For this study, we downloaded the light curves from the Q1–Q17 \textit{Kepler} Data Release 25 (DR25), made available by the \textit{Kepler} mission through the Mikulski Archive for Space Telescopes. We processed the downloaded light curves with the pre-processing method described in Section 2.2.
 
We also used the set of Kepler Objects of Interest (KOIs) available on the NASA Exoplanet Archive as of 3 June 2023. There are 9,564 total dispositioned KOIs with DV light curves available. These include 2,350 confirmed planets, 2,366 candidates, and 4,848 false positives. Transit parameters (period, epoch, duration) taken from the KIC catalog are used for testing and verification. We downloaded all 9564 KOIs in the Kepler Input Catalog (KIC).

As a start, we chose a subset of the KOIs which contains only target stars whose transit events are all labeled as "CONFIRMED" planetary transits. This filters out all light curves whose target star contains a transit event with \textit{Kepler} disposition of "FALSE POSITIVE" or "CANDIDATE". This process is needed to ensure that none of the target stars we include are affected by undesirable false positive interference, or contain secondary eclipses of eclipsing binaries. The metadata for secondary eclipses are not recorded in the KOIs and therefore cannot be fully masked when needed.

From this KOI subset, we prepared two sets of light curves to test the GPFC system: one \rt{transit} data set and one \rt{non-transit} data set. The \rt{transit} data set was composed of light curves of all target stars that contain transits of a confirmed USP planet. For each such star, the light curve was conditioned by keeping the USP transits intact while removing the transits from any other planets. The \rt{non-transit} data set consisted of light curves of all target stars with all of their transit events masked.
 
Through this process, we obtained 43 light curves with verifiable USP transits for the \rt{transit} dataset, and 1,437 light curves devoid of any transit events for the \rt{non-transit} dataset.

\subsection{Recovery of All Confirmed \textit{Kepler} Ultra-Short-Period Exoplanets}

By applying the GPFC method to the \rt{transit} dataset, it has been demonstrated that GPFC accurately identifies all of the 43 confirmed USPs. The method recovers these exoplanets at the periods within 0.004\%  of the recorded value in the KOIs as shown in \rt{Figure \ref{fig:real_usp_yt_p1} and Figure \ref{fig:real_usp_yt_p2} in the Appendix section.} Additionally, all confirmed USPs are recovered with a score of 0.99 or higher, further reinforcing the validity of the GPFC method. This crucial validation lays the foundation for using GPFC to uncover new exoplanets in the \textit{Kepler} database, which will be elaborated in our follow-up paper (Wang et al. 2024, submitted).

\begin{figure}
  \centering
  \includegraphics[width=0.48\textwidth]{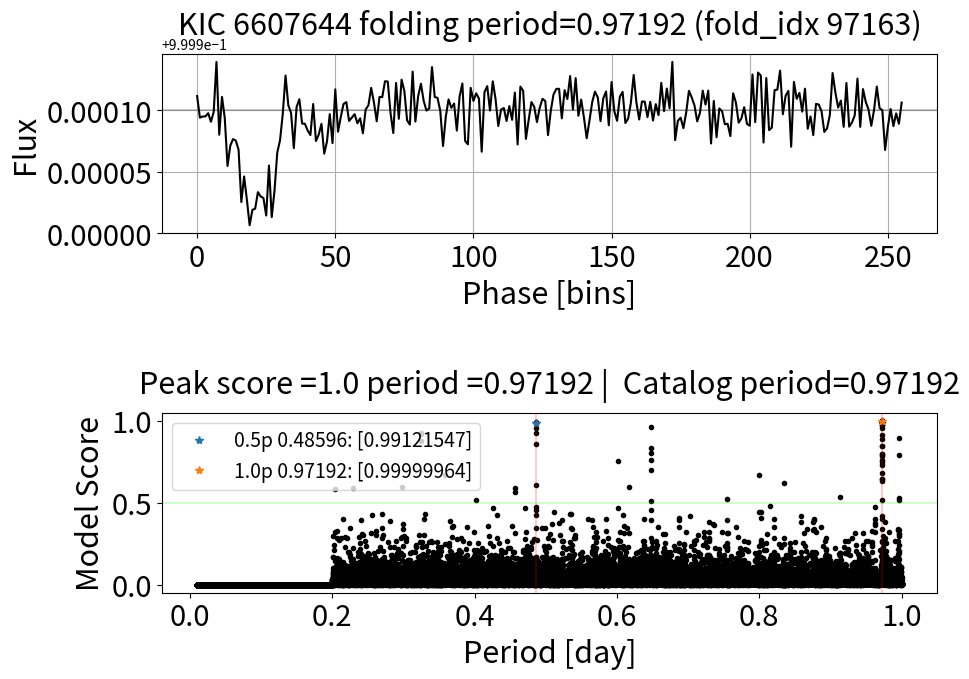}
    \caption{An example of a real \textit{Kepler} USP (KID 6607644) recovered by the GPFC method. The phase fold at the detected transit period of 0.971923 days is shown in the upper plot. In tandem, the model score vs. trial period plot (the lower plot) shows a clear peak at period 0.971923 days with model score of 0.99999964. This peak period accurately matches the known transit period of 0.971916 days in the \textit{Kepler} catalog (error $7.362 \cdot 10^{-6}$). The model scores also reveal corresponding peaks at the harmonics of the transit period. For instance, at the half period 0.48596 days the model reaches a high score of 0.99121547.}
    \label{fig:real_usp_yt_illustrate}
\end{figure}

Fig.~\ref{fig:real_usp_yt_illustrate} demonstrates an example USP that the GPFC method recovered with a high probability score.

\begin{figure}
  \centering
  
  \begin{minipage}[c]{0.32\textwidth}
    \centering
    \subfloat{\includegraphics[width=\textwidth]{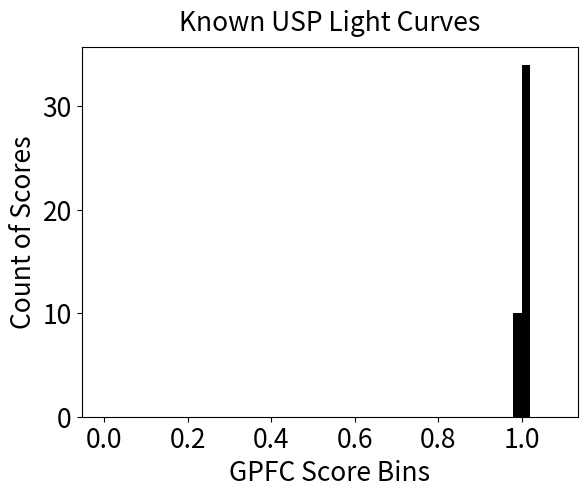}}\label{fig:dist_real_usp_yt}
  \end{minipage}
  
  \vspace{10pt}
  
  \begin{minipage}[c]{0.32\textwidth}
    \centering
    \subfloat{\includegraphics[width=\textwidth]{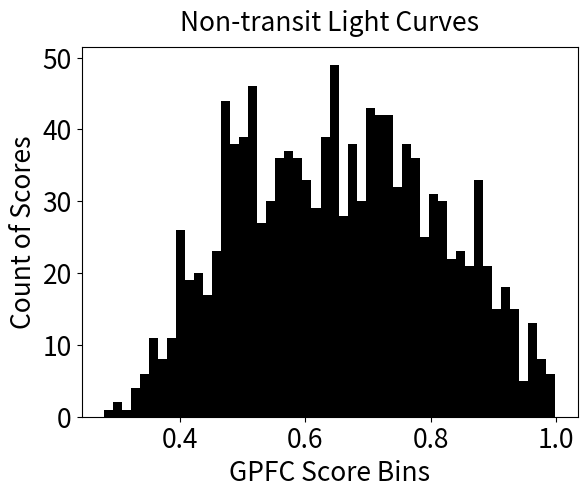}}\label{fig:dist_resl_usp_nt}
  \end{minipage}

  \caption{The performance of GPFC on real \textit{Kepler} light curves. As shown in the upper panel, GPFC distinguishes all confirmed USPs with scores above 0.99. The lower panel represents GPFC's scores on real \textit{Kepler} KOI with known transits masked. We use the \textit{Kepler} KOI with only confirmed planets to avoid the secondary eclipses of eclipsing binaries. \rt{For the 1,437 \textit{Kepler} KOIs with known transits masked, signals that received high scores were further evaluated as potential candidates. We set a cut-off CNN score threshold of 0.8 for this test, aligning with a 96\% true-positive rate (TPR) as indicated in our simulation test.}}
  \label{fig:dist_real_usp_scores}
\end{figure}

In Fig.~\ref{fig:dist_real_usp_scores}, the upper plot shows the distribution of the GPFC model scores for the 43 confirmed USPs, and the lower plot shows the score distribution for all of the \rt{non-transit} light curves. The score for each light curve is determined by identifying the optimal peak among the 100,000 scores that exceed a height of $3\sigma$ and have a width of more than 3 data points. We see that most of the \rt{non-transit} light curves received scores below 0.9. The candidates on \rt{non-transit} light curves which the model identified with scores above 0.8 identified in \rt{non-transit} light curves will be further examined in our follow-up investigation. \rt{Our simulation results indicated that CNN scores of 0.76, 0.8, and 0.9 correspond to true-positive rates (TPR) of 90\%, 96\%, and 98\%, respectively. Based on these findings, we selected 0.8 as the threshold for this test. For the validation of real candidates, we implemented a pipeline comprising a series of rigorous verification tests. This validation procedure led to the identification of five promising candidates, which will be elaborately discussed in our upcoming follow-up paper.}

\begin{figure*}
    \centering
    \includegraphics[width=1.0\textwidth]{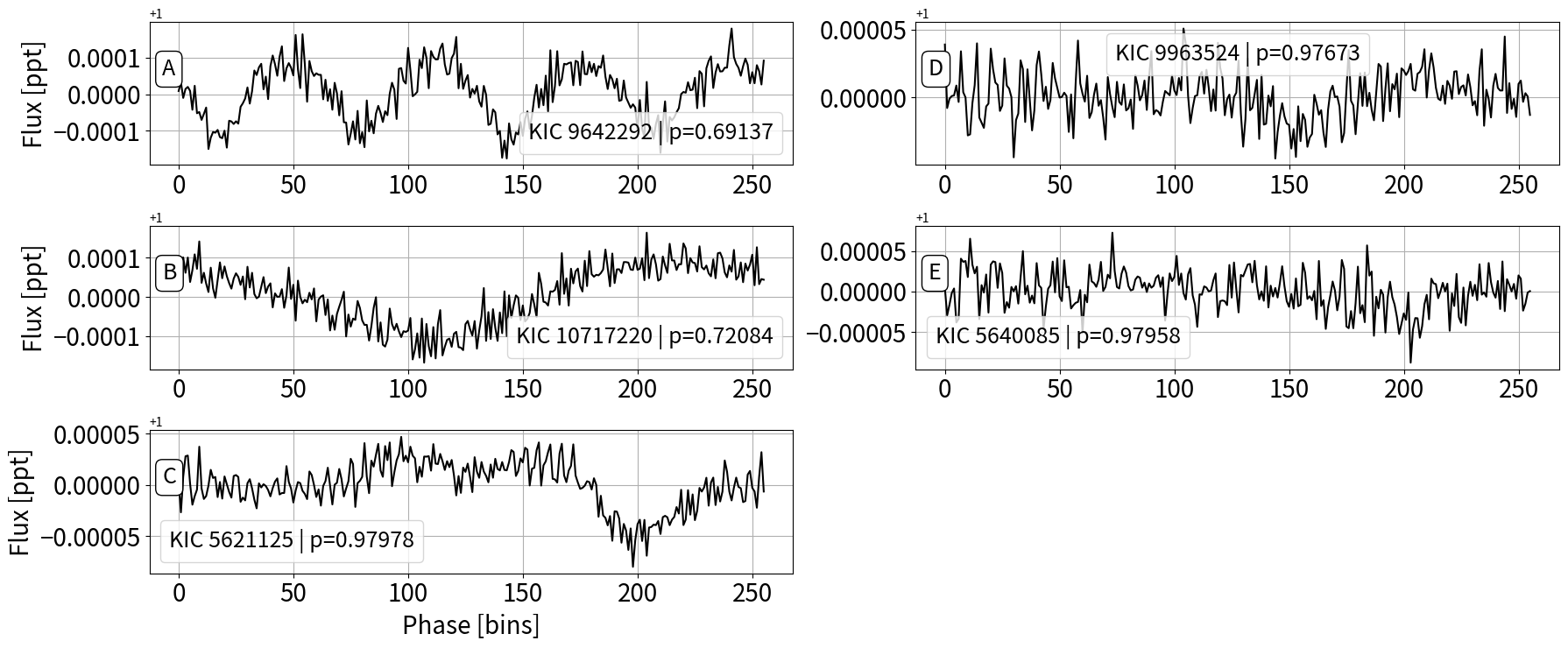}
    \caption{High-probability (greater than 0.9) false positives identified by the GPFC method. The GPFC method identified various types of high-probability false positives. Plot A illustrates an example of false positives caused by sinusoidal signals, while plot B presents those due to longer-term stellar variations. These can be effectively filtered out in an enhanced pre-processing process. Plot C exhibits noise from extraneous sources. Conversely, plot D and E display transit-shaped signals; however, due to the noisy nature of the light curves, the detected transit-like signals require further False Alarm Probability check in our automated verification process. The specifics of this verification will be detailed in our forthcoming discovery paper (Wang et al. 2024, submitted).}
    \label{fig:real_usp_nt_false_positive}
\end{figure*}

\rt{Fig.~\ref{fig:real_usp_nt_false_positive} illustrates the high-probability (probability greater than 0.9) false positives detected by the GPFC method.} \rt{In our analysis of 1,437 Kepler light curves, we identified several false positives: 10 due to sinusoidal signals, exemplified by KIC 9642292 in plot A, and 32 resulting from longer-term stellar variation, as illustrated by KIC 10717220 in plot B. Additionally, two cases showed significant noise from unrelated sources, represented by KIC 5621125 in plot C. These 44 false positives were manually vetted and subsequently discarded. Looking ahead, we anticipate that such false positives could be automatically filtered out in future work through improved pre-processing procedures, eliminating the need for manual intervention.}

\rt{In contrast, plot D (KIC 9963524) and plot E (KIC 5640085) present transit-shaped signals; however, given the noisy nature of the light curves, these detected signals are subjected to further False Alarm Probability (FAP) checks in the verification process. This process, which operates without human intervention, aims to eliminate most of the false positives originating from extraneous sources (such as exemplified in plot D and E). The details of this verification process will be discussed in our forthcoming discovery paper.}

\section{Discussion and Conclusion}

\subsection{Real Exoplanet Discovery}

For the purpose of validating the GPFC approach with actual \textit{Kepler} light curves, we confined our work to target stars exclusively associated with confirmed planetary transit events. As a subsequent step, we intend to employ GPFC for real exoplanet discovery, encompassing the entire \textit{Kepler} catalog.

The design of the GPFC system is inherently tailored to detect shallow and narrow transit signals, combining high-precision phase folding with deep learning techniques. Consequently, it holds potential for uncovering smaller exoplanets. Our preliminary runs of GPFC on real \textit{Kepler} survey data indicate its capability to quickly process vast data sets. Moreover, it has exhibited proficiency in identifying exoplanet transit signals with lower SNRs, which might have previously gone unnoticed. Our team plans to continue exploring this avenue and will provide updates on our findings in an incoming publication (Wang et al. 2024, submitted).

\subsection{GPFC vs the fBLS method}

\citet{Shahaf_2022} introduced a fast model of BLS, termed fBLS, which incorporates a fast folding algorithm and a transit detection approach optimized for run time. This fBLS system reported the discovery of six candidates: KIC 6293500, 9217391, 9835433, 2718885, 6359893, and 11187332. To further validate the accuracy of GPFC, we applied GPFC to these six candidates and all of them were accurately identified with periods precisely matching those reported by fBLS. Both the GPFC and fBLS methods have similar run times of approximately 6 seconds to analyze a typical light curve comprising 65536 data points.

In the folding phase, GPFC is designed to directly process the original \textit{Kepler} light curves, which are irregularly sampled with sporadic time gaps. This is achieved by folding based on the original measurement time modulo the trial period. In contrast, due to the requirement of uniform sampling by the fast-folding algorithm (FFA), the fBLS approach needs to initiate its analysis with a "brute-force" folding procedure to address the irregular samples. Concurrently, the fBLS algorithm folds the light curve using approximately 250,000 trial periods within the interval [0.2, 1.0] days. Our simulation tests, as evidenced by the ROC curve, indicated that GPFC, when folded with 250k periods, exhibited performance similar to that observed when folding with 100k periods (AUC values are 0.901 and 0.902, respectively) , so we opted to retain our configuration at 100k trial periods.

During the transit detection phase, GPFC uses 256 flux bins to represent the folded light curve in contrast to the 40 to 80 bins used by fBLS. Whereas fBLS and other optimized BLS algorithms iterate through a pre-selected set of transit start and duration combinations, the GPFC method differentiates itself by incorporating a deep learning module (CNN), to pinpoint the transit at any location and with any duration. Given that the light curves used to train the CNN incorporate randomized transit phases and durations, the CNN naturally recognizes an expansive spectrum of transits without compromising processing speed.

\subsection{GPFC vs a GPU-BLS method}

We noted the presence of a GPU-accelerated BLS software on GitHub, termed \href{https://johnh2o2.github.io/cuvarbase/bls.html}{cuvarbase BLS}. Cuvarbase BLS is configured to automatically assign trial frequencies and features a default regular running mode. Moreover, it provides an additional $use\_fast$ mode designed for expedited processing, albeit with somewhat reduced functionality. We conducted a performance comparison between GPFC and cuvarbase BLS, using a typical \textit{Kepler} light curve comprising 70k data points. In our evaluation, cuvarbase BLS processed the data in 30 seconds using its default mode and 1.05 seconds with the $use\_fast$ mode. In contrast, GPFC accomplished the same task in 6 seconds. The speeds of GPFC and GPU-BLS are comparable, \rt{while GPFC demonstrates higher accuracy, with AUC 7\% higher than that of GPU-BLS, which is consistent with comparisons to the traditional BLS method.}

\subsection{Further Performance Boosting}

Although the GPFC method is notably faster than the traditional BLS method, there is still room to optimize its speed without compromising accuracy. As indicated in Fig.~\ref{fig:accuracy-auc-cnn-vs-bls}, the performance enhancement of GPFC starts to plateau at around 60k trial periods. In our research, we consistently used GPFC with 100k periods because our speed is not an issue in most cases. However, for specific scenarios, like aiming to search through the entire \textit{Kepler} catalog in less than 10 days, achieving a speed faster than the current 6 seconds might be beneficial. In such contexts, by switching to 60k trial periods, the computation time could be reduced to approximately 2.5 seconds. 

Another approach to enhance the speed of the GPU Phase Folding is to further parallelize the algorithm. Due to the memory constraints of our GPU, which has a capacity of only 11 GB, we are limited to evaluating a maximum of 16 trial periods. By leveraging a GPU with a larger memory capacity or one supporting higher parallelism for atomic read-write operations, the run-time of our GPFC system will be improved even further.

\subsection{Extending the Applications of GPFC}

While our primary focus in this research has been on USP exoplanets within the \textit{Kepler} survey, the GPFC method is inherently generic. The GPU phase folding algorithm can function with any set of trial periods, Moreover, the CNN is amenable to training on any custom-configured dataset. \rt{Consequently, the GPFC method is readily adaptable for searching transits across a wider range of periods. Within the period range of 1 to 10 days, the application of the GPFC method is expected to be straightforward and similar to current practices. For period ranges that exceed 10 days, the grid search space significantly widens, and it would be beneficial to employ additional strategies to preliminarily identify potential regions of interest.}

\rt{For longer periods, we can conduct a Kepler catalog analysis similar to that used for the ultra-short-period, to examine the distribution of planet radii and transit durations. This allows us to ascertain the required $N_{sample}$ for Equation \ref{sampling_equation}. For larger planets characterized by longer transit durations, $N_{sample}$ can be smaller, suggesting the viability of a less precise search grid. Such an adjustment is anticipated to decrease the runtime, as depicted in Figure \ref{fig:bls-speed-upto-100k}. Larger planets usually produce bigger transit signals with higher signal-to-noise ratios (SNR), whereas longer periods result in a reduced number of folds in the light curve, subsequently lowering the noise reduction effect achieved through phase folding. Further predictions of our method's performance across these varying scenarios can be made by a thorough analysis of the Kepler database in the target period ranges.}

\rt{Furthermore, another natural progression would be to enhance the final CNN model by incorporating additional metadata, as demonstrated in various CNN models \citep{malik2022discovering}. This augmentation is expected to boost performance and potentially reduce the occurrence of false positives.}

Moreover, the adaptability of GPFC extends to light curves from a range of other transit surveys, including K2, TESS, and upcoming missions like Plato (as detailed in \citep{Plato_Rauer_2022, Plato2_Rauer_2014}) and the Earth 2.0 (ET) mission (\citep{GeJ22_Earth2, GeJ22_Earth2_wp, GeJ22_Earth2_mission}), and beyond. \rt{We expect that the pre-processing step of the GPFC method will need to be tailored for each specific survey to suit the unique characteristics of its light curves. Apart from this customization, the GPFC method is broadly applicable across different surveys.}

\subsection{Conclusion}

This paper introduces the GPFC transit signal detection method, examining its ability to detect periodic transits in stellar photometric time series. GPFC innovatively merges a GPU phase folding algorithm with a Convolutional Neural Network, aiming to identify transit signals from small exoplanets directly from original raw light curves without resorting to the traditional BLS transit detection method.

Using simulated light curves based on \textit{Kepler} parameter distributions, we compared the accuracy and performance of the GPFC method against the BLS method. Our focus is primarily on scenarios with low SNRs and we see that for SNR < 10, GPFC has an advantage in performance over BLS.

In terms of computational speed, GPFC holds a significant advantage. To match the performance levels set by GPFC, BLS would need to utilize at least 20,000 frequencies. This requirement makes BLS exponentially slower — approximately three orders of magnitude — compared to GPFC. When operating with 100,000 frequencies and using a fast version of BLS provided by AstroPy, accelerated by Cython, BLS is still roughly 15 times slower than GPFC.

Lastly, when the GPFC method was applied to known \textit{Kepler} exoplanets, it successfully recovered all confirmed planets within the period search range of [0.2, 1.0] days, as covered in this paper. It also assigned low scores to \textit{Kepler} light curves where all transits were masked. During our tests, GPFC consistently identified new, small transit signals in \textit{Kepler} light curves rapidly, showing its potential applications in large-scale transit surveys.

\subsection{Acknowledgements}
JG acknowledges the support from the Strategic Priority Program on Space Science of Chinese Academy of Sciences under grant No. XDA15020600. This research has made use of NASA's Astrophysics Data System and the NASA Exoplanet Archive, operated by the California Institute of Technology, under contract with NASA under the Exoplanet Exploration Program. This paper includes data collected by the \textit{Kepler} mission. Funding for the \textit{Kepler} mission is provided by the NASA Science Mission directorate. 

\section*{Data Availability}
The \textit{Kepler} light curves used in this study can be accessed from \url{https://exoplanetarchive.ipac.caltech.edu/cgi-bin/TblView/nph-tblView?app=ExoTbls&config=cumulative}. A table with all confirmed physically periodic sources will be provided online.

\bibliography{refs}
\bibliographystyle{mnras}

\appendix
\section{Additional Figures}

\begin{figure*}
    \centering
    \includegraphics[width=1.0\textwidth]{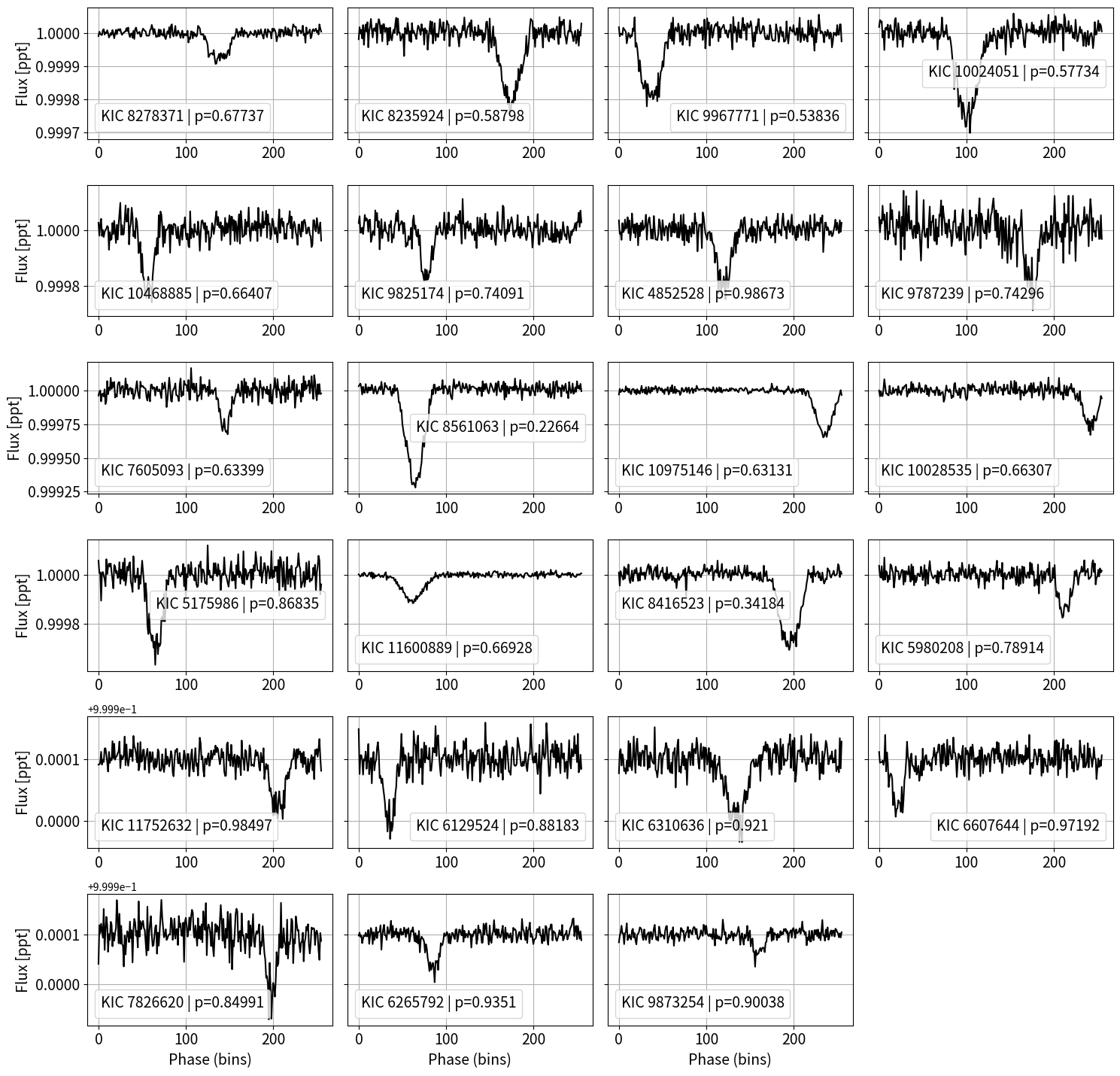}
    \caption{GPFC recovery of confirmed USPs. The GPFC method identified all of the 43 confirmed USPs in \textit{Kepler} with each of them given a CNN score above 0.99. All of the detected periods are the exactly same as the period recorded in the KIC catalog. Each folded light curve exhibits a clear transit signal. This blind search test proved the validity of the GPFC method in terms of USP transit detection. }
    \label{fig:real_usp_yt_p1}
\end{figure*}

\begin{figure*}
    \centering
    \includegraphics[width=1.0\textwidth]{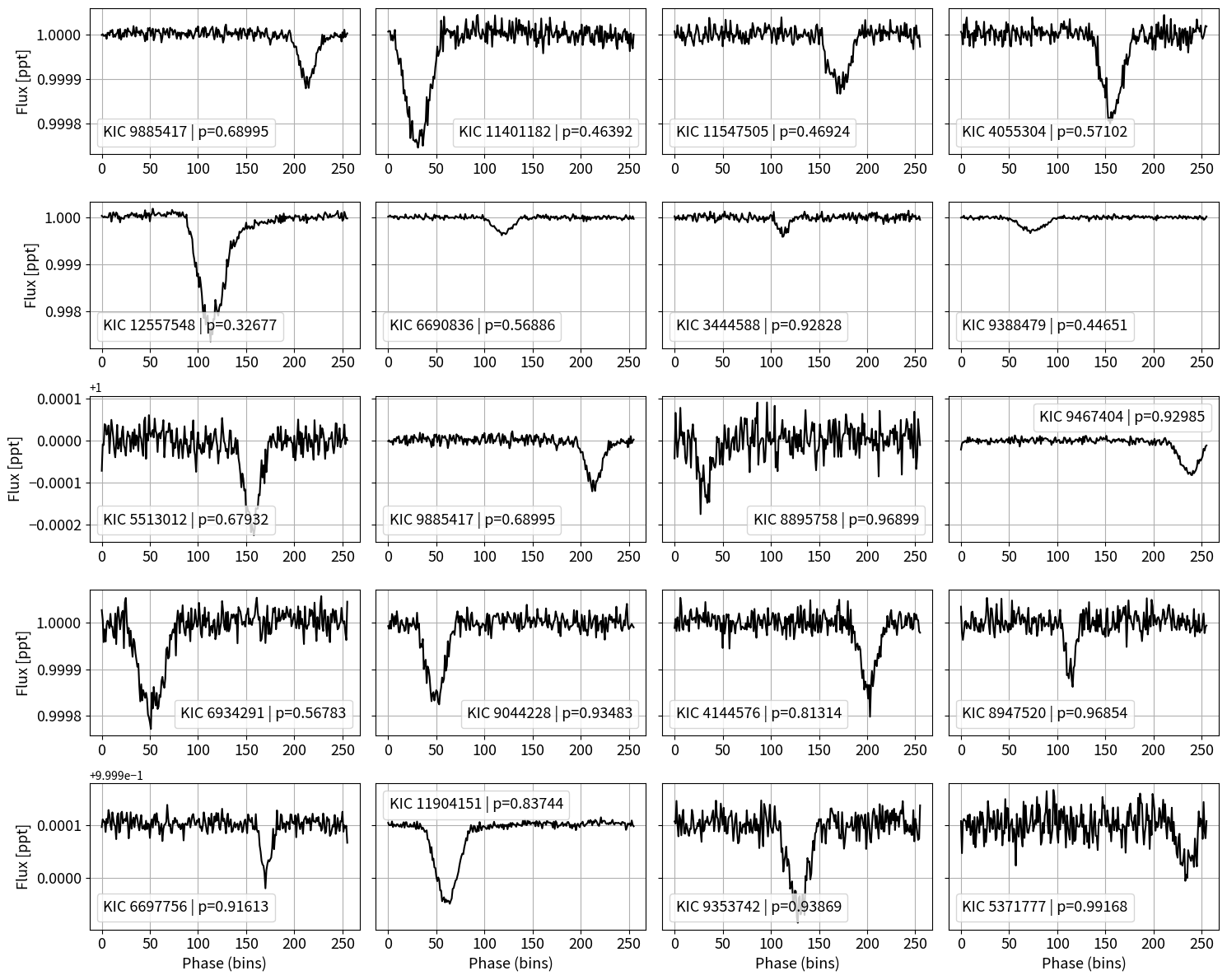}
    \caption{(Continued) GPFC recovery of confirmed USPs. The GPFC method identified all of the 43 confirmed USPs in \textit{Kepler} with each of them given a CNN score above 0.99. All of the detected periods are the exactly same as the period recorded in the KIC catalog. Each folded light curve exhibits a clear transit signal. This blind search test proved the validity of the GPFC method in terms of USP transit detection. }
    \label{fig:real_usp_yt_p2}
\end{figure*}

\bsp	
\label{lastpage}

\end{document}